\documentclass[12pt]{article}
\newlength{\abstractwidth}
\renewcommand{\thanks}[1]{\footnote{#1}} 

\newcommand{\be}{\begin{equation}}
\newcommand{\bea}{\begin{eqnarray}}
\newcommand{\eea}{\end{eqnarray}}
\newcommand{\ee}{\end{equation}}
\newcommand{\N}{{\cal N}}
\newcommand{\<}{\langle}
\renewcommand{\>}{\rangle}
\def\ba{\begin{eqnarray}}
\def\ea{\end{eqnarray}}

\def\N{{\cal N}}

\def\O{{\cal O}}

\def\Im{{\rm Im}}

\def\det{{\rm det}}
\def\half{ {1\over 2}}

\def\14{{1 \over 4}}
\def\p{\partial}
\def\tet{\vartheta}
\def\no{\nonumber}
\def\e{\epsilon}

\begin{document}
\baselineskip=16pt

\begin{flushright}
UCLA/04/TEP/47 \\
Columbia/Math/04 \\
2004 November 19
\end{flushright}

\bigskip

\begin{center}
{\Large \bf  ASYZYGIES, MODULAR FORMS, AND
\\
\medskip
THE SUPERSTRING MEASURE II
\footnote{Research supported in
part by National Science Foundation grants PHY-01-40151 and DMS-02-45371.}}

\bigskip\bigskip

{\large  Eric D'Hoker$^a$ and D.H. Phong$^b$} \\ 

\bigskip

$^a$ \sl Department of Physics and Astronomy \\
\sl University of California, Los Angeles, CA 90095 \\

\smallskip

$^b$ \sl Department of Mathematics \\ 
\sl Columbia University, New York, NY 10027

\end{center}

\bigskip\bigskip

\begin{abstract}

Precise factorization constraints are formulated for the three-loop superstring chiral measure, in the separating degeneration limit. Several natural  Ans\"atze in terms of polynomials in theta constants for the density of the measure are examined. None of these Ans\"atze turns out to satisfy the dual criteria of modular covariance of weight 6, and of tending to the desired degeneration limit. However, an Ansatz is found which does satisfies these criteria for the square of the density
of the measure, raising the possibility that
it is not the density of the measure, but its square
which is a polynomial in theta constants. A key notion is that of
totally asyzygous sextets of spin structures. It is argued that the Ansatz produces a vanishing cosmological constant.

\end{abstract}


\newpage

\baselineskip=15pt
\setcounter{equation}{0}
\setcounter{footnote}{0}
\newtheorem{theorem}{Theorem}
\newtheorem{definition}{Definition}

\vfill\eject

\section{Introduction}
\setcounter{equation}{0}

Recently, the superstring measure to two loop order and for even
spin structure was computed from first principles \cite{I,II,III,IV,V,dp02,adp}.
The construction relies on a careful treatment of supermoduli, chiral
splitting and finite-dimensional gauge fixing determinants, and builds
on earlier work in this direction \cite{dp88, dp89}. Although intermediate 
calculations are complex and intricate, the final form of the superstring
measure turns out to be very simply expressed in terms of a new 
modular object, denoted by $\Xi _6 [\delta] (\Omega)$ in \cite{IV}.

\medskip

At present, no analogous derivation is available to 3-loop order and beyond.
Some of the special simplicity of genus 2 does carry over to genus 3,
in that no Schottky relations need to be imposed on the period matrix.
The structure of supermoduli, however, becomes considerably more
complex and, at present, the calculation appears formidable.

\medskip

Therefore, the simplicity of the ultimate form of the two-loop superstring measure
raises the question as to whether the genus 3 superstring measure might
have a comparatively simple form in terms of natural modular objects.
Constraints from holomorphicity, modular invariance, and physical
factorization will provide powerful restrictions on any candidate measures. 
The precise form of the 2-loop measure gives a drastic constraint on the 
separating degeneration limits of the  3-loop measure.\footnote{The 
constraints of modular invariance were used along these lines to 
guess the bosonic string measure to 2- and 3-loops in 
\cite{moore} and \cite{bkmp} respectively. A general theory based on
constraints from modular invariance and physical factorization was 
developed in \cite{deg}.}

\medskip

In this paper, we take a first step in the degeneration approach to the 
superstring measure by formulating a precise Ansatz for
the 3-loop measure and verifying that it satisfies the correct
factorization conditions when the worldsheet degenerates.
Our Ansatz for the (chiral) superstring measure $d\mu[\Delta](\Omega^{(3)})$ 
can be described as follows. Set
\be
\label{Ansatz}
d\mu[\Delta](\Omega^{(3)})
=
{\tet[\Delta](0,\Omega^{(3)})^4
\,
\Xi_6[\Delta](\Omega^{(3)})
\over
8\pi^4\,\Psi_9(\Omega^{(3)})}
\
\prod_{I\leq J}d\Omega_{IJ}^{(3)}
\ee 
Here $\Delta$ is a fixed even spin structure, 
$\Omega^{(3)}=\{\Omega_{IJ}^{(3)}\}$ is the period matrix of the  genus 3 worldsheet,
$\Psi_9(\Omega_{IJ}^{(3)})^2$ is the modular form
$\Psi_{18}(\Omega^{(3)})$ of weight 18 constructed in
\cite{igusa}, and the measure 
$\Psi_9(\Omega_{IJ}^{(3)})^{-1}\prod d\Omega_{IJ}$
has been shown to be holomorphic in \cite{bkmp}. The key term $\Xi_6[\Delta](\Omega^{(3)})$
is to be determined by the following constraints:

\medskip

({\it i}) $\Xi_6[\Delta](\Omega^{(3)})$ is holomorphic in 
$\Omega^{(3)}$ on the Siegel upper half space;

\medskip

({\it ii}) $\Xi_6[\Delta](\Omega^{(3)})$ is a modular covariant form 
of weight 6 in the sense that, under modular transformations sending $\Omega_{IJ}^{(3)}\to\tilde\Omega_{IJ}^{(3)}=
(A\Omega^{(3)}+B)(C\Omega^{(3)}+D)^{-1}$, $\Delta\to\tilde\Delta$, we have
\be
\Xi_6[\tilde\Delta](\tilde\Omega^{(3)})
=
\e(\Delta,M)^4
\det\,(C\Omega^{(3)}+D)^6
\,
\Xi_6[\Delta](\Omega^{(3)}),
\ee
where $\e(\Delta,M)$ is the same phase factor as in the modular 
transformation for $\tet$-constants.

\medskip

({\it iii}) In the degeneration $t\to 0$,
where the worldsheet separates into
a genus $1$ and a genus $2$ surface of period matrices $\Omega^{(1)}$ and $\Omega^{(2)}$ respectively, we must have
\be
\lim_{t\to 0} \Xi_6[\Delta](\Omega^{(3)})
=
\eta(\Omega^{(1)})^{12}
\
\Xi_6[\delta](\Omega^{(2)}),
\ee
where $\Xi_6[\delta](\Omega^{(2)})$ is the main new factor in the genus $2$ superstring measure found in \cite{I,IV}. 

\medskip

The constraint ({\it iii}) on the degeneration limit of  $\Xi_6[\Delta](\Omega^{(3)})$
is a consequence of the factorization properties of string amplitudes. 
To establish it, we require a precise formula for the degeneration 
limit of the measure $\Psi_9(\Omega^{(3)})^{-1}\prod d\Omega_{IJ}^{(3)}$, 
formula which is also one of the main results of this paper 
(see Theorem 1 below).

\medskip
We should stress that the condition ({\it iii}) is very restrictive, since it applies to an {\it arbitrary} separating degeneration. Thus we have to expect $\Xi_6[\Delta](\Omega^{(3)})$ to be built of sums of many terms,
different groups of which would tend to 0 in different limits.

\medskip

The original expression for $\Xi_6[\delta](\Omega^{(2)})$ derived in \cite{I,II,III,IV} depended very much on the fact that the worldsheet had genus $2$. Since then, two alternate expressions have been found which can extend to higher genus \cite{dp04}.
A characterizing feature of these two expressions is that one
of them is a sum over fourth powers of $\tet$-constants
of {\it triplets} of spin structures,
while the other is a sum of second powers of
$\tet$-constants of {\it sextets} of spin structures.  
The key to determining which $N$-tuplets $\{\delta_i\}$ of spin structures should contribute to $\Xi_6[\delta](\Omega)$ turns out to be the notion of total asyzygies. Recall that
to any triplet of spin structures $\{\delta_1,\delta_2,\delta_3\}$ is associated a modular invariant sign, namely the product 
\bea
e(\delta _1, \delta _2, \delta_3)=\<\delta_1|\delta_2\>\,\<\delta_2|\delta_3\>
\<\delta_3|\delta_1\>
\eea
of relative signatures 
$\< \delta |\epsilon \> = \exp  4 \pi i (\delta ' \epsilon '' - \epsilon ' \delta '')$. 
A triplet of spin structures is said to be syzygous or asyzygous, depending on whether $e$  is $+1$ or $-1$. The criteria for which triplets or sextets should contribute to $\Xi_6[\delta](\Omega)$ turns out to be entirely
expressible in terms of asyzygies
(see \S 5.1 below). Once the criteria for which triplets or sextets to include has been identified, one needs to find
phase assinments $\epsilon(\delta;\{\delta_i\})$ with which to
sum the contributions of various sextets. The phase assignments have to be consistent with modular invariance, which identifies them all up to a global phase.

\medskip

These alternative descriptions of $\Xi_6[\delta](\Omega)$ suggest
several possible generalizations to genus 3, all involving
summations over monomials in $\tet[\Delta_i]$. They are listed in
\S 5.2, where we describe also in detail their viability as Ans\"atze 
for the genus 3 superstring chiral measure
$\Xi_6[\Delta](\Omega^{(3)})$. The net outcome is the following:

\medskip

$\bullet$ A first Ansatz is in terms of sums of products of three fourth powers 
only, such as $\tet [\Delta_{i_1}]^4 \tet [\Delta_{i_2}]^4\tet [\Delta_{i_3}]^4$.
Using in particular the degeneration formulas
of \cite{dp04}, we show that none of this form exists which
satisfies the criteria ({\it ii}) and ({\it iii}).
More generally,  the criterion ({\it iii}), requiring the appearance of $\eta(\Omega^{(1)})$, effectively prevents the rule for which $N$-tuplets to
be included to remain the same for all genera.  

\medskip

$\bullet$ Next, we consider Ans\"atze involving sums of second powers, 
such as $\prod_{j=1}^6\tet[\Delta_{i_j}]^2$.
In genus $2$, the sextets which contribute to $\Xi_6[\delta](\Omega)$ 
can be characterized by the condition
of $\delta$-admissibility (see \S 5.1). This condition makes sense for 
all genera, but in genus $3$, the set of such sextets
(called $\Delta$-admissible by extension) breaks up into many orbits 
under the subgroup of modular transformations fixing a given spin 
structure $\Delta$. One particularly important orbit is the set of sextets 
which do not contain $\Delta$, and which are {\it totally asyzygous},
in the sense that any of their sub-triplets is asyzygous.
We refer to the other orbits as {\it partially asyzygous}.
The partially asyzygous orbits do not appear to have as simple a description as the orbit of totally asyzygous sextets,
although they can be identified by computer analysis.

\medskip
The partially asyzygous orbits turn out not to be viable candidates for $\Xi_6[\Delta](\Omega^{(3)})$: computer analysis reveals that many 
of them do not admit consistent phase
assignments $\epsilon(\Delta;\{\Delta_i\})$. Even when they do,
their degeneration limits do not satisfy the criterion ({\it iii}) listed above. 
Thus we rule them out as Ans\"atze for
$\Xi_6[\Delta](\Omega^{(3)})$.

\medskip
We found the criterion of totally asyzygous sextets to be much more compelling: its key property is that the genus $2$ sextets
obtained by factorization from a totally asyzygous genus 3 sextet
automatically satisfy the key condition of admissibility in genus 2 (see Lemma 1 in section \S 6.1). Furthermore, although these genus 2 sextets may be admissible but not $\delta$-admissible, Lemma 2 in section \S 6.2 shows that the contributions of such sextets sum up to 0 if they are assigned phases consistent with modular invariance. Thus the Ansatz in terms of totally asyzygous sextets would satisfy the degenerating condition ({\it iii}) if phase assignments exist which are consistent with ({\it ii}). 
However, perhaps surprisingly, such a consistent phase assignment
does not exist and ({\it ii}) cannot be satisfied. A simple example is provided in section \S 6.2.2.

\medskip

$\bullet$ Another possible Ansatz could be in terms of sums of  
products of twelve  first powers of $\tet$, such as 
$\prod_{i=1}^{12}\tet[\Delta_i](0,\Omega^{(3)})$.
The criterion for which dozens $\{\Delta_i\}_{1\leq i\leq 12}$
to include is difficult to guess from the genus 2 case.
There is no consistent phase assignments if the dozens are
assumed to consist of a pair of totally asyzygous sextets,
and more generally, no consistent sign assignments appear possible.

\medskip
Thus, we are led to believe that no candidate for 
$\Xi _6 [\Delta] (\Omega ^{(3)})$ exists which is a polynomial in $\tet$.
On the other hand, consistent modular covariant assignments 
$\e(\Delta;\{\Delta_i\},\{\Delta_i'\})$ do exist for suitable bilinear 
combinations of pairs of totally asyzygous
sextets of $\tet [\Delta _i]^2$. 
This suggests that only $\Xi_6[\Delta](\Omega^{(3)})^2$ is a 
polynomial in $\tet$-constants. 
We find that, for a suitable integer normalization factor $N$,
and a suitable choice of multiplicities $N_{pq}$ of the orbits 
${\cal Q}_{pq}$ of pairs $\{\Delta_i,\Delta_i'\}$ of totally asyzygous sextets under the subgroup of $Sp(6,{\bf Z})$ fixing $\Delta$, the expression
\be
\label{Ansatz1}
\Xi_6[\Delta](\Omega^{(3)})^2
=
{1\over 2^8N}\sum_{pq}N_{pq}
\sum_{(\{\Delta_i\},\{\Delta_i'\})\in{\cal Q}_{pq}}
\e(\Delta;\{\Delta_i\},\{\Delta_i'\})
\prod_{i=1}^6\tet^2[\Delta_i]
\prod_{i=1}^6\tet^2[\Delta_i']
\ee
does satisfy all the conditions implied by ({\it i})-({\it iii}) for the square of $\Xi_6[\Delta](\Omega^{(3)})$, for arbitrary separating degeneration limits. In particular, it is a highly non-trivial 
result that in any separating degeneration limit of this form to 
a genus 2 and a genus 1 surface, the limit becomes a perfect square.
In general, these expressions will not admit holomorphic square roots
away from the separating degeneration limit. 
If there exists  a specific choice of multiplicities $N_{pq}$ (not all $0$) 
which  guarantees the existence of a holomorphic square root, then
(\ref{Ansatz1}) will single out a compelling candidate for  the genus $3$  
superstring measure. The existence of such a holomorphic square 
root is known to occur at genus 3 in at least one other instance,
namely the modular form $\Psi _9 (\Omega ^{(3)})
= \prod _\Delta \tet [\Delta] (0,\Omega ^{(3)}) ^\half$, which is known
to be the (unexpectedly) holomorphic square root of $\Psi _{18} (\Omega ^{(3)})$.

 \medskip
 
The remainder of this paper is organized as follows.
In section 2, the general criterion for physical factorization is spelled out
for the superstring measure. In section 3, the factorization properties
of the bosonic factors in the genus 3 measure are derived.
In section 4, the construction of the genus 3 superstring measure is
formulated as a degeneration problem. In section 5, the consistency
with criteria {\it (i), (ii), (iii)} above  of various candidates is analyzed
and (\ref{Ansatz1}) is constructed.

\newpage

\section{Factorization of the superstring measure}
\setcounter{equation}{0}

The main goal of this section is to derive the
precise degeneration constraints which the 3-loop superstring measure
must satisfy when a separating cycle in the worldsheet $\Sigma^{(3)}$ is pinched to a point, and $\Sigma^{(3)}$ separates into a torus $\Sigma^{(1)}$ and a genus $2$ surface
$\Sigma^{(2)}$.

\subsection{Geometric picture of factorization}

We begin with the geometric description of the moduli space of
Riemann surfaces near the divisor of surfaces with nodes, as provided by the following well-known construction \cite{fay}.

\medskip

Let $\Sigma^{(1)}$ and $\Sigma^{(2)}$ be two Riemann surfaces
of genus $h_1$ and $h_2$, let $p_1\in \Sigma^{(1)}$, $p_2\in \Sigma^{(2)}$ be two given points, and let $|z_1|<1$,
$|z_2|<1$ be local coordinates on $\Sigma^{(1)}$ and $\Sigma^{(2)}$ which are centered at $p_1$ and $p_2$
respectively. Let ${\cal S}$ be the surface given by
${\cal S}=\{(X,Y,t);\ XY=t\ ,|X|<1,\,|Y|<1,\,|t|<1\}$, and construct the
fibration ${\cal C}$ of surfaces over the unit disk $\{t;|t|<1\}$
given by
\be
{\cal C}=\{(z_1,t);z_1\in\Sigma^{(1)},\ |z_1|>|t|\}
\ \cup\
{\cal S}
\ \cup\
\{(z_2,t);z_1\in\Sigma^{(2)},\ |z_2|>|t|\},
\ee
with the following identifications
\bea
&&
(z_1,t)\sim (z_1,{t\over z_1},t)\ \ {\rm for}\ z_1\in\Sigma^{(1)},
\ |t|<|z_1|<1
\nonumber\\
&&
(z_2,t)\sim ({t\over z_2},z_2,t)\ \ {\rm for}\ z_2\in\Sigma^{(2)}, \ |t|<|z_2|<1.
\eea
For each $t\not=0$, the fiber of ${\cal S}$ above $t$
can be identified with the annulus
$A_t=\{X;|t|<|X|<1\}$. Thus 
the fiber of ${\cal C}$ above $t$
is a regular surface $\Sigma_t$ of genus $h=h_1+h_2$,
which can be covered by the three overlapping charts $\Sigma^{(1)}\setminus
\{|z_1|>|t|\}$, $A_t$, and $\Sigma^{(2)}\setminus \{|z_2|>|t|\}$, with the identifications
\be
z_1\,\sim\, X\,\sim\, {t\over z_2},
\ \ \
{\rm for}\ \ |t|<|z_1|,|z_2|<1.
\ee

\subsection{Physical picture of factorization}

In the physical picture, we view the surface $\Sigma_t$ rather as the disjoint union
\be
\Sigma_t
=
\Sigma_{in}^{(1)}
\ 
\cup
\
A_t
\
\cup
\Sigma_{out}^{(2)}
\ee
where we have set $\Sigma_{in}^{(1)}=\Sigma^{(1)}\setminus\{|z_1|<1\}$,
and 
$\Sigma_{out}^{(2)}=\Sigma^{(1)}\setminus\{|z_1|<1\}$.
In a given conformal field theory, the surfaces with boundary $\Sigma_{in}^{(1)}$,
$\Sigma_{out}^{(2)}$ define two states $\<\Sigma_{in}^{(1)}|$
and $|\Sigma_{out}^{(2)}\>$. To make contact with the Hamiltonian
picture, we can use the exponential map
$\xi\to X=t^{\half}e^\xi$ to identify the annulus $A_t$ with a cylinder
\be
\{\xi=\xi_0+i\xi_1;\ 0\leq\xi_1\leq 2\pi,
\ -\half\ln{1\over|t|}<\xi_0<\half\ln{1\over|t|}\}.
\ee
Now the operators for time and space translations are the
Hamiltonian $H=L_0+\bar L_0$ and the momentum operator
$P=L_0-\bar L_0$
\footnote{Since all conformal anomalies ultimately cancel,
we can ignore the contribution of the central charge
when we map the annulus into the cylinder.}.
If we view $\xi_0$ as ``time", and $\xi_1$ as ``space", 
then the shift in time and the shift in space corresponding
to the cylinder are given respectively by the length of the cylinder and the phase shift in $\xi$ as the point $X$
moves on a straight line from $X=|t|$ to $X=1$.
This gives
$-\ln |t|$ for the shift in time
and $\arg (t) $ for the shift in space, since
$\xi=|t|^{\half}e^{-i\half\arg(t)}$
and $\xi=|t|^{-\half}e^{i\half\arg(t)}$ are the points
on the cylinder corresponding to $X=|t|^{\half}$
and $X=1$. The cylinder corresponds then to the following
operator insertion
\be
\exp \bigg (i\arg(t)(L_0-\bar L_0) \bigg )
\,
\exp \bigg (\ln (|t|) (L_0+\bar L_0) \bigg )
=
t^{L_0}\,\bar t^{\bar L_0}
\ee
and hence the partition function ${\cal Z}_t$ corresponding to the surface
$\Sigma_t$ is given by
\be
{\cal Z}_t
=
\<\Sigma_{in}^{(1)}|\ t^{L_0}\,\bar t^{\bar L_0}\ |\Sigma_{out}^{(2)}\>
\ee
To obtain the degenerating limit $t\to 0$, we insert a
basis of states $|\psi_\alpha\>$
diagonalizing $t^{L_0}\,\bar t^{\bar L_0}$
\be
{\cal Z}_t
=
\sum_\alpha\ \<\Sigma_{in}^{(1)}|\psi_\alpha\>\,
\<\psi_\alpha|\ t^{L_0}\,\bar t^{\bar L_0}\ |\psi_\alpha\>
\,
\<\psi_\alpha|\Sigma_{out}^{(2)}\>
\ee
The descendant states $|\psi_\alpha\>$ contribute lower order terms in the limit $t\to 0$. To identify the leading contribution, we need thus to consider only primary states.
In the case of string propagation, before the GSO projection,
the state with lowest $m^2$
is the tachyon with $m^2=-2$. By momentum conservation,
its momentum must be $k^\mu=0$ (it is not on-shell, but intermediate states do not have to be on-shell). Since the vertex
for tachyon emission with momentum $0$ is just the identity,
the leading term for ${\cal Z}_t$ is given by
\be
{\cal Z}_t
=
{\cal Z}^{(1)}\cdot t^{-2}\bar t^{-2}\cdot
{\cal Z}^{(2)}
+
O(|t|^{-3})
\ee
where ${\cal Z}^{(1)}$ and ${\cal Z}^{(2)}$ are the partition functions for the surfaces $\Sigma^{(1)}$ and $\Sigma^{(2)}$.

\medskip
To deal with spin structures, we start from surfaces $\Sigma^{(i)}$ with canonical homology bases $A_I^{(i)}$, $B_I^{(i)}$, $\#(A_I^{(i)}\cap B_J^{(i)})=\delta_{IJ}$,
$\#(A_I^{(i)}\cap A_J^{(i)})=0$, $\#(B_I^{(i)}\cap B_J^{(i)})=0$
for $1\leq I,J\leq h_i$. Then the combined bases give a canonical
basis for the genus $h_1+h_2$ surface $\Sigma_t$. With this choice of homology bases,
a spin structure $\Delta$ can be identified with an assignment
of either $0$ or $1/2$ to each homology cycle of $\Sigma_t$,
and hence with a pair $(\delta_1,\delta_2)$, with $\delta_i$ a spin structure on the surface $\Sigma^{(i)}$
\be
\Delta=\pmatrix{\delta_2\cr\delta_1}.
\ee
In a conformal field theory where the fields are world sheet fermions requiring
a spin structure, the preceding degeneration formula becomes
\be
{\cal Z}_t[\Delta]
=
{\cal Z}^{(1)}[\delta_1] \cdot t^{-2}\bar t^{-2}\cdot
{\cal Z}^{(2)}[\delta_2]
+
\O(|t|^{-3}).
\ee

\subsection{Factorization of the genus $3$ superstring measure}

We formulate now the precise degeneration constraint for the superstring 
measure when the worldsheet $\Sigma=\Sigma_t$ is of genus $h=3$ and 
degenerates into two surfaces $\Sigma^{(1)}$ and $\Sigma^{(2)}$
of genus $h_1=1$ and $h_2=2$.

\medskip
We shall assume that, at loop order $h$, the vacuum-to-vacuum superstring amplitude is of the form
\be
\label{fullintegral}
{\cal A}
=
\sum_{\Delta,\bar\Delta}
c_{\Delta,\bar\Delta}\int_{{\cal M}_h}(\det\, \Im\,\Omega^{(h)})^{-5}
\
d\mu[\Delta](\Omega^{(h)})
\,\wedge
\,
\overline{d\mu[\bar\Delta](\Omega^{(h)})}
\ee
where $c_{\Delta,\bar\Delta}$ are suitable phases, and the sum
over the spin structures $\Delta,\bar\Delta$ corresponds to the GSO projection, which projects out the tachyon and produces space-time supersymmetry. The space ${\cal M}_h$ is the moduli space of Riemann surfaces of genus $h$. 
We always fix a homology basis, and view each Riemann surface as 
characterized by its 
period matrix $\Omega^{(h)}=\{\Omega_{IJ}^{(h)}\}_{1\leq I,J\leq h}$.
The form $d\mu[\Delta](\Omega)$ is a $(3h-3,0)$ holomorphic form
on ${\cal M}_h$, transforming under modular transformations
in such a way that the full expression above is modular invariant. It is called the (chiral) superstring measure at
genus $h$.

\medskip
Near $t=0$, the $3h-3$ moduli parametrizing $\Sigma_t$ can be chosen to be the $3h_1-3$ and $3h_2-3$ moduli for the surfaces
$\Sigma^{(1)}$ and $\Sigma^{(2)}$, together with the 3 parameters
$p_1,p_2$ and $t$. The degeneration formulas derived above for
conformal field theory suggest imposing the following degeneration
constraint for the chiral superstring measure
\be
\label{degeneration}
d\mu[\Delta](\Omega)
=
d\mu[\delta_1](\Omega^{(1)}) \, \wedge \, {dt\over t^2} \,
\wedge \, d\mu[\delta_2](\Omega^{(2)}) \, \wedge \, dp_1\wedge dp_2
+ \O(t^{-1})
\ee
As usual, these formulas hold for $h_1,h_2\geq 2$. When $h_1=1$,
the counting is slightly different, since $p_1$ and its differential are no longer relevant due to translation invariance on the torus. This is actually the case of main interest in the
present paper, so we make the above formula more explicit in this
case: the moduli for $\Sigma^{(1)}$ is then a single parameter
$\Omega^{(1)}$, and the superstring measure for one-loop
is $\tet^4[\delta_1](\Omega^{(1)})/2^5\pi^4\eta^{12}(\Omega^{(1)})$
(see e.g. \cite{IV}, eq. (8.2)). Thus the degeneration constraint for
the chiral superstring measure at genus $h$ when the worldsheet
separates into a torus $\Sigma^{(1)}$ and a genus $h-1$ surface $\Sigma^{(2)}$ is given by
\be
\label{superdeg}
d\mu[\Delta](\Omega)
=
{\tet^4[\delta_1](\Omega^{(1)})
\over
2^5\pi^4\eta^{12}(\Omega^{(1)})}\,d\Omega^{(1)} \, \wedge \, {dt\over t^2}
\, \wedge \, d\mu[\delta_2](\Omega^{(2)}) \, \wedge \, dp_2 + \O(t^{-1}).
\ee

\subsection{Factorization of the genus $3$ bosonic string measure}

Although this paper is mainly concerned with the genus $3$ superstring measure and its degeneration limit, we take the opportunity to discuss also similar issues for the bosonic
string, partly as a check later on our method. The measure for the bosonic string in the critical dimension is of the form
\be
{\cal A}=\int_{{\cal M}_h}(\det\, \Im\,\Omega)^{-13}\
d\mu_B(\Omega)\,\wedge\,\overline{d\mu_B(\Omega)}
\ee 
where $d\mu_B(\Omega)$ is holomorphic. Because the intermediate state of lowest mass is still the tachyon, the measure
$d\mu_B(\Omega)$ satisfies the same
degeneration constraint as in (\ref{degeneration}).
When the worldsheet $\Sigma$ degenerates into a torus $\Sigma^{(1)}$ and a surface of genus $2$,
the degeneration constraint can be written as
\be
\label{bosonic}
d\mu_B(\Omega)
=
{d\Omega^{(1)}
\over
(2\pi)^{12}\eta^{24}(\Omega^{(1)})}  \, \wedge \, {dt\over t^2} \,
\wedge \, d\mu_B(\Omega^{(2)}) \, \wedge \, dp_2 + \O(t^{-1})
\ee
where $(2\pi)^{-12}\eta^{-24}(\Omega^{(1)})\,d\Omega^{(1)}$ is the genus $1$ bosonic string measure,
with the conventions of \cite{dp88} and the normalization
$d^2\Omega^{(1)}/(8\pi^2\, \Im\,\Omega^{(1)})^2$
for the $SL(2,{\bf R})$ invariant measure on the Siegel upper half space.

\newpage

\section{The measure 
$\prod_{I\leq J}d\Omega_{IJ}^{(3)}/\Psi_9(\Omega^{(3)})$ in genus $3$}
\setcounter{equation}{0}

An important feature of the chiral superstring measure
$d\mu[\Delta](\Omega^{(h)})$ is that it is a holomorphic $(3h-3,0)$ form. 
To find it, we begin by constructing a natural holomorphic $(3h-3,0)$ form $d\mu_B(\Omega^{(h)})$ on ${\cal M}_h$
(later identified with the chiral bosonic measure, but this is not essential for our considerations), so that the problem of finding $d\mu[\Delta](\Omega^{(h)})$ 
reduces to that of finding
the density $d\mu[\Delta]/d\mu_B$. In genera $h=2$ and $h=3$,
we can exploit the fact that ${\cal M}_h$ and the Siegel upper
half space of symmetric matrices with positive imaginary part
have the same dimension, and henceforth we consider only these
cases. 

\subsection{The modular forms $\Psi_{18}(\Omega^{(3)})$ and $\Psi_{10}(\Omega^{(2)})$}

Recall that on a surface $\Sigma$ of genus $h$, there are
$2^{2h}$ spin structures, of which $2^{h-1}(2^h+1)$ are even
and $2^{h-1}(2^h-1)$ are odd. The parity of a spin structure
$\Delta$ corresponds to the parity in $\zeta$ of the $\vartheta$-function $\tet[\Delta](\zeta,\Omega^{(h)})$,
which is also the parity of the number of independent
holomorphic spinors of spin structure $\Delta$. The properties
of $\tet$-functions which we need can be found in \cite{IV},
\S 2.1-\S 2.3 and \cite{adp}, Appendix B. For convenience, we 
restate here the transformations of spin structures $\Delta\to
\tilde\Delta$ and $\tet$-constants $\tet[\Delta](0,\Omega^{(h)})\to
\tet[\tilde\Delta](0,\tilde\Omega^{(h)})$ under modular 
transformations
\be
\label{modulartransformation}
\tilde\Omega^{(h)}=(A\Omega^{(h)}+B)(C\Omega^{(h)}+D)^{-1},
\qquad M=\pmatrix{A &B\cr C& D\cr}\in Sp(2h,{\bf Z}).
\ee
If we write $\Delta=(\Delta'|\Delta'')$
and $\tilde\Delta=(\tilde\Delta'|\tilde\Delta'')$,
they are given by
\be
\pmatrix{\tilde\Delta'\cr\tilde\Delta''}
=
\pmatrix{D&-C\cr -B&A\cr}\pmatrix{\Delta'\cr\Delta''}
+
{1\over 2}
{\rm diag}\,\pmatrix{CD^T\cr AB^T}
\ee
and by
\be
\tet[\tilde\Delta](0,\tilde\Omega)
=
\epsilon(\Delta,M)
\,
\det(C\Omega^{(h)}+D)^{\half}
\,
\tet[\Delta](0,\Omega^{(h)}),
\ee
where $\epsilon(\Delta,M)$ is an eighth root of unity, which depends 
on both the spin structure $\Delta$ and the modular
transformation $M$. There is no simple closed formula for
$\epsilon(\Delta,M)$, but its values for $h=2$ on generators of 
$Sp(4,{\bf Z})$ can be found in \cite{IV}, \S 2.3. 

\medskip

The above transformation for $\tet$-constants should be compared 
with the defining transformation law for modular forms 
$\Phi(\Omega)$ of a given weight $w$
\be
\Phi(\tilde\Omega^{(h)})
=
\det(C\Omega^{(h)}+D)^w\,\Phi(\Omega^{(h)})
\ee
which do not involve roots of unity
such as $\epsilon(\Delta,M)$. Nevertheless,
the following natural form can be defined using the even $\tet$-constants
\be
\label{Psi}
\Psi_{2^{h-1}(2^h+1)k}(\Omega^{(h)})
=
\prod_{\Delta\ even}\tet^{2k}[\Delta](0,\Omega^{(h)})
\ee
It has been shown by Igusa \cite{igusa} that in genus $h=2$ and $h=3$, 
$\Psi_{2^{h-1}(2^h+1)k}(\Omega^{(h)})$ are modular forms of weight 
$2^{h-1}(2^h+1)k$ when $k=1$ and $k=1/2$ respectively. 

\medskip

Let these forms be denoted by $\Psi_{10}(\Omega^{(2)})$ and $\Psi_{18}(\Omega^{(3)})$ respectively. It is well-known that the form
$\Psi_{10}(\Omega^{(2)})$ has no zero inside the moduli space of
Riemann surfaces of genus $2$, while the form $\Psi_{18}(\Omega^{(3)})$ 
vanishes exactly of second order along
the variety of hyperelliptic surfaces of genus $3$
\cite{bkmp}. Indeed, $\Psi_{2^{h-1}(2^h+1)}(\Omega^{(h)})$ vanishes
if and only if a $\tet$-constant vanishes for some even spin
structure $\Delta$. Since the parity of the number of independent
holomorphic spinors is the same as the parity of $\Delta$,
this means that there are at least $2$ independent holomorphic
spinors of spin structure $\Delta$. By the Riemann-Roch theorem,
the number of zeroes of a holomorphic spinor is always $(h-1)$.
In genus $h=2$, a holomorphic spinor has then exactly one zero,
and the ratio of two linearly independent holomorphic spinors
would be a meromorphic function with exactly one zero and one pole. 
Such a function provides a one-to-one correspondence between the 
given Riemann surface and the sphere, contradicting
our initial assumption that $h=2$. Similarly, when $h=3$,
a holomorphic spinor has $2$ zeroes, and the ratio of two linearly 
independent holomorphic spinors is a meromorphic function with 
two zeroes and two poles. Such a function provides a two-to-one 
correspondence with the sphere, and thus the Riemann
surface must be hyperelliptic. 
Conversely, if $s^2=\prod_{i=1}^8(x-u_i)$ is a hyperelliptic surface of 
genus $3$, then $s^{-\half}(dx)^{\half}$ and $xs^{-\half}(dx)^{\half}$ 
define two holomorphic spinors associated with an even spin structure. 
Thus $\Psi_{18}(\Omega^{(3)})$ vanishes at such surfaces (in fact,
to second order), and the proof of the claim is complete.

\medskip
Since the form $\Psi_{18}(\Omega^{(3)})$ vanishes of second order, we can follow \cite{bkmp} and obtain a holomorphic character $\Psi_9(\Omega^{(3)})$ by taking its square root
\be
\Psi_9(\Omega^{(3)})^2
=
\Psi_{18}(\Omega^{(3)})
\ee
 
\medskip
In genus $h=2$ and $h=3$, the moduli space ${\cal M}_h$ and the Siegel upper half space have the same dimension, which is 3 and 6
respectively. An integral over ${\cal M}_h$ can be identified with an integral over a fundamental domain of the modular group
$Sp(2h,{\bf Z})$ in the Siegel upper half space. On this space,
we can introduce the following holomorphic $(3h-3,0)$ forms
\footnote{The ordering of the forms $d\Omega_{IJ}^{(h)}$ in these measures is a matter of convention. We shall ignore the resulting
$\pm$ signs and sometimes denote the resulting volume form just by $d^{h(h+1)/2}\Omega^{(h)}$.}
\bea
\label{volumeform}
&&{1
\over
\Psi_{10}(\Omega^{(2)})}
\,
\prod_{1\leq I\leq J\leq 2}d\Omega_{IJ}^{(2)},
\ \ {\rm for\ genus}\ h=2
\nonumber\\
&&{1
\over
\Psi_{9}(\Omega^{(3)})}
\ \
\prod_{1\leq I\leq J\leq 3}d\Omega_{IJ}^{(3)},
\ \ {\rm for\ genus}\ h=3.
\eea
Both measures are holomorphic on the Siegel upper half space.
This is obvious when $h=2$. When $h=3$, this is due to \cite{bkmp},
who showed that the form $\prod_{I\leq J}d\Omega_{IJ}^{(3)}$ 
also vanishes along the variety of hyperelliptic surfaces,
so that the zeroes in the denominator $\Psi_{9}(\Omega^{(3)})$
are cancelled by the measure factor.

\medskip

It follows from Igusa's classification theorem for genus $2$ modular 
forms that the bosonic string measure is actually
given in genus $h=2$ by \cite{bkmp,moore}
\be
d\mu_B(\Omega^{(2)})=
\ {c_2 \over \Psi_{10}(\Omega^{(2)})}
\,
\prod_{1\leq I\leq J\leq 2}d\Omega_{IJ}^{(2)},
\ee
where $c_2$ is an overall constant. This constant was in fact evaluated 
in \cite{IV} \S 7.1, and was found to be $c_2 = \pi ^{-12}$.
There is no such classification theorem in genus $3$ or higher,
but cogent arguments have been proposed
for the similar relation in genus $3$ to hold
\cite{bkmp}
\be
d\mu_B(\Omega^{(3)})
=
\ {c_3 \over  \Psi_{9}(\Omega^{(3)})}
\ \
\prod_{1\leq I\leq J\leq 3}d\Omega_{IJ}^{(3)}
\ee
with $c_3$ another overall constant.
As part of our program for
determining the genus $3$ superstring measure,
we shall present further evidence for this relation below.

\subsection{Degeneration of $\Psi_{9}^{-1}(\Omega^{(3)})\prod_{I\leq J}d\Omega_{IJ}^{(3)}$}

The superstring chiral measure will be identified by its density
with respect to the basic measure $\Psi_9(\Omega^{(3)})\prod_{I\leq J}d\Omega_{IJ}^{(3)}$.
In order to reformulate the degeneration constraints
(\ref{degeneration}) for the superstring measure in terms
of degeneration constraints for its density, we need the precise
degeneration limit of the measure $\Psi_9(\Omega^{(3)})\prod_{I\leq J}d\Omega_{IJ}^{(3)}$.
This is given in the following theorem:

\begin{theorem}
In the degeneration limit given by \S 2.1, $t\to 0$, we have
\bea 
{1 \over \Psi_{9}(\Omega^{(3)})}
\ \
\prod_{1\leq I\leq J\leq 3}d\Omega_{IJ}^{(3)}
= { 1 \over (2 \pi )^6 }
 {\prod_{I\leq J}\Omega_{IJ}^{(2)}
 \over
 \Psi _{10} (\Omega^{(2)})}
  \wedge 
  {d\Omega^{(1)} \over  \eta (\Omega^{(1)})^{24}}
\wedge {dt \over t^2} \wedge dp_2
+\O({1\over t}).
\eea
\end{theorem}

\medskip

\noindent
{\it Proof.} We consider the parametrization of surfaces $\Sigma_t$ 
degenerating into two surfaces $\Sigma^{(1)}$ and $\Sigma^{(2)}$ 
described in \S 2.1. As indicated there, we choose canonical homology bases $(A_{I_i}^{(i)},B_{I_i}^{(i)})$, so that the union of these
cycles constitutes a canonical homology basis for $\Sigma$.
Let $(\omega_{I_1}^t,\omega_{I_2}^t)$ be the basis of 
holomorphic Abelian differentials on $\Sigma_t$ dual to
the $(A_{I_1},A_{I_2})$ cycles. Then these holomorphic differentials have the following asymptotic behavior as $t\to 0$
\cite{fay}
\bea
\omega ^t _{I_1} (z)
& = & \left \{ \matrix{
\omega _{I_1} (z) + { t \over 4} \omega _{I_1} (p_1) \omega ^{(1)} _{p_1} (z)
+ \O (t^2)
& {\rm when} & z \in \Sigma^{(1)} \cr
 { t \over 4} \omega _{I_1} (p_1) \omega ^{(2)} _{p_2} (z)
+ \O (t^2)
& {\rm when} &  z \in \Sigma^{(2)} \cr } \right .
\no \\ && \\
\omega ^t _{I_2} (z)
& = & \left \{ \matrix{
 { t \over 4} \omega _{I_2} (p_2) \omega ^{(1)} _{p_1} (z)
+ \O (t^2)
& {\rm when} & z \in \Sigma^{(1)} \cr
\omega _{I_2} (z) + { t \over 4} \omega _{I_2} (p_2) \omega ^{(2)} _{p_2} (z)
+ \O (t^2)
& {\rm when} &  z \in \Sigma^{(2)} \cr } \right .
\no
\eea
Here, $\omega ^{(i)} _{p_a}$ refers to the meromorphic differential
on surface $\Sigma^{(i)}$, $i=1,2$ with a double pole at $p_i$,
while $\omega _{I_i}$  refers to a basis of
holomorphic differentials on surface $\Sigma^{(i)}$.

\medskip

The components of the period matrix behave as follows
\cite{fay},
\bea
\Omega ^t _{I_1 J_1}
=
\Omega ^{(1)} _{I_1 J_1} +
{i \pi \over 2} t \omega _{I_1} (p_1) \omega _{J_1} (p_1)
& \hskip .5in &
\Omega ^t _{I_1 J_2}
=
 {i \pi \over 2} t \omega _{I_1} (p_1) \omega _{J_2} (p_2)
\no \\
\Omega ^t _{I_2 J_2}
=
\Omega ^{(2)} _{I_2 J_2} +
{i \pi \over 2} t \omega _{I_2} (p_2) \omega _{J_2} (p_2)
& \hskip .5in &
\Omega ^t _{I_2 J_1}
=
 {i \pi \over 2} t \omega _{I_2} (p_2) \omega _{J_1} (p_1)
\eea
where $\Omega ^{(i)}$ refers to the period matrix on the surface $\Sigma^{(i)}$.

\bigskip
Henceforth, we consider the case where $h_1=1$ and $h_2=2$.
It is convenient to set $\Omega^{(1)}=\tau$,
$\Omega^{(2)}=\Omega$, and  
use the following notations,
\bea
\Omega ^{(3)} = \left ( \matrix{
\Omega _{11} & \Omega _{12} & \tau_1 \cr
\Omega _{12} & \Omega _{22} & \tau_2 \cr
\tau _1 & \tau _2 & \tau_3 \cr} \right )
\hskip 1in
\left \{ \matrix{
\tau _1 & = & {i\pi\over 2} t ~ \omega _1 (p_2) \omega _0 (p_1) \cr
\tau _2 & = & {i\pi\over 2} t ~ \omega _2 (p_2) \omega _0 (p_1) \cr} \right .
\eea
Here $\omega _I(p_2)$ denote the genus 2
holomorphic differentials and $\omega _0$ denotes the genus
1 holomorphic differential which is just the constant $1$
in the usual parametrization of the torus of modulus $\tau$
as ${\bf C}/{\bf Z}+\tau{\bf Z}$, since the homology
basis $(A_1^{(1)},B_1^{(1)})$ has been fixed.

\medskip

The $\tet$-constants at genus 3 for even spin structures
behave differently in the separating limit depending on
whether the spin structures on the genus 2 and genus 1
components are both even or both odd. We have the
following limits,
\bea
\label{thetaconst}
\tet \left [ \matrix{\delta \cr \mu \cr} \right ] (0, \Omega ^{(3)})
& = &
\tet [ \delta ] (0, \Omega ) ~ \tet [\mu ] (0, \tau) + \O (t)
\no \\
\tet \left [ \matrix{\nu \cr \nu_0 \cr} \right ] (0, \Omega ^{(3)})
& = &
{ t \over 4} \omega _0 (p_1) \tet ' _1 (0, \tau) h_\nu (p_2)^2 + \O (t^2)
\eea
Here, $\delta$ (resp. $\nu$) denote an even (resp. odd) genus 2
spin structure, while $\mu$ denotes an even genus 1 spin structure
and $\nu_0$ denotes the unique genus 1 odd spin structure.
Furthermore, we use the familiar notation,
\bea
h_\nu (z)^2  \equiv \omega _I (z) \p ^I \tet [\nu ] (0, \Omega)
\eea
This square is defined for any surface, while its square
root, $h_\nu$ in single-valued only on a surface with
spin structure $\nu$.

\medskip

\noindent
{\sl a) The limit of $\Psi _{18}$}

\medskip

We are now in a position to study the limit of the modular
form $\Psi _{18}$ and its square root $\Psi _9$.
In genus $3$, there are 36 even spin structures, of which
30 separate into two even spin structures in genus
$1$ and $2$, and 6 separate into two odd spin structures
in genus $1$ and $2$. In the first group of 30, the spin 
structures obtained after degeneration run over all 10 
genus $2$ even spin structures and over all 3 genus 1 even spin structures.
Similarly, in the second group of 6, the spin structures
obtained after degeneration run over all 6 genus 2 even spin structures. 
Thus we obtain 
\bea
\Psi _{18} (\Omega ^{(3)})
= 
\prod _{\delta, \mu} \left ( \tet [\delta ](0, \Omega ) \tet [\mu ](0,\tau) \right )
~
\prod _\nu \left ( {t \over 4} h_\nu (p_2)^2 \omega _0 (p_1) \tet _1 ' (0,\tau) \right )
+ \O (t^7)
\eea
In view of the well-known genus $1$ identities,
\be
\label{genus1identity}
\tet '_1 (0,\tau) = -2 \pi \eta (\tau)^3,
\qquad
\prod _\mu \tet [\mu ](0, \tau) = 2 \eta (\tau)^3,
\ee
and the definition of $\Psi_{10}(\Omega)$,
this can be rewritten as
\bea
\Psi_{18}(\Omega^{(3)})
& = &
\Psi _{10} (\Omega) ^{3/2} \left ( 2 \eta (\tau)^3 \right )^{10}
\left ( {t \over 4} \right )^6 \omega _0 (p_1)^6 \left ( - 2 \pi \eta (\tau)^3 \right )^6
\prod _\nu h_\nu (p_2)^2 + \O (t^7)
\no \\
& = & 2^4 \pi ^6 t^6 \omega _0 (p_1) ^6 \Psi _{10}(\Omega) ^{3/2}
\eta (\tau)^{48} \prod _\nu h_\nu (p_2)^2 + \O (t^7)
\eea
Taking the square root, we find
\bea
\label{Psi}
\Psi _9(\Omega^{(3)}) = 4 \pi^3 t^3 \omega _0(p_1)^3 \Psi _{10} (\Omega )^{3/4}
\eta (\tau)^{24} \prod _\nu h_\nu (p_2) + \O (t^4)
\eea
Notice that, while each $h_\nu$ may not be single-valued on a
surface with given spin structure (or without specified spin structures),
the product over all $\nu$ is single-valued on any surface.

\medskip

\noindent
{\sl b) The limit of the volume factor $d^6 \Omega ^{(3)}_{IJ}$}

\medskip

We turn next to the limit of the measure $d^6 \Omega ^{(3)}_{IJ}$.
In the above notation, we have
\bea
\label{dOmega}
d^6 \Omega ^{(3)} _{IJ}
=
d^3 \Omega \wedge d\tau \wedge d\tau_1 \wedge d \tau _2
\eea
We now evaluate $d \tau_1 \wedge d \tau_2$, using the definition
of its ingredients,
\bea
d \tau_1 \wedge d \tau_2
=
- {\pi ^2 \over 4} ~ t dt \wedge dp_2 ~ \omega _0 (p_1)^2
\bigg ( \omega _1 (p_2) \p \omega _2 (p_2)
        - \omega _2 (p_2) \p \omega _1 (p_2) \bigg )
\eea
The combination in parentheses is a holomorphic 3-form in $p_2$.
To evaluate it, we turn to the hyperelliptic representation
of Riemann surfaces of genus $2$. Let the surface $\Sigma^{(2)}$ be given by
\be
s^2=\prod_{i=1}^6(x-u_i)
\ee
Then $z^{J-1}dz/s(z)$, $J=1,2$, is a basis of holomorphic differential forms. Let $\sigma_{IJ}$ be the change of bases matrix from this basis to the basis $\omega_{I_2}^{(2)}(z)$
(which we abbreviate to $\omega_I(z)$ for the rest of the proof of Theorem 1),
\bea
\label{sigma}
2 \pi i\, \omega _I (z) = \sum _J \sigma _{IJ} { z^{J-1} dz \over s(z)}
\eea
Hence, we have
\bea
\omega _1 (p_2) \p \omega _2 (p_2)
        - \omega _2 (p_2) \p \omega _1 (p_2)
=
- { 1 \over 4 \pi^2} (\det \sigma ) { (dp_2)^3 \over s(p_2)^2}
\eea
Thus the holomorphic 3-form manifestly has
6 simple zeros precisely at the branch points, exactly as
$\prod _\nu h_\nu (z)$. Thus, the $p_2$-dependence
of these two forms is the same. 

\medskip

\noindent
{\sl c) Determining the constant of proportionality}

\medskip

We need to determine
the constant of proportionality, which is moduli dependent,
and requires several precise coefficients of proportionality
between the $\tet$-function and the hyperelliptic representation
of holomorphic spinors \cite{IV}. 
In the hyperelliptic representation, each of the 6 odd spin
structures $\nu_i$ corresponds to a branch point $u_i$,
and the one-form $h_{\nu_i}^2(z)$
is proportional to the one-form $(x-u_i)dz/s(z)$. Set
\be
\label{proportionality}
h_{\nu_i}^2(z)=\N_{\nu_i}(x-u_i){dz\over s(z)}
\ee 
where $\N_{\nu_i}$ is a moduli dependent constant.
Then we have
\bea
{ (dp_2)^3 \over s(p_2)^2} 
=
\left ( \prod _i { 1 \over \N _i ^{1/2}} \right ) \prod _\nu h_\nu (p_2)
\eea
Combining all, we obtain
\bea
\label{dtau1}
d \tau_1 \wedge d \tau_2
=
{1 \over 16} ~ t dt \wedge dp_2 ~ \omega _0 (p_1)^2
{\det \sigma \over \prod _i \N _i ^{1/2}} \prod _\nu h_\nu (p_2)
\eea
Next, we have the following two  identities
\bea
\label{2identities}
(\det \sigma ) ^4 \tet [\delta ]^8 
&= &
\prod _{i<j} (a_i-a_j)^2 (b_i-b_j)^2
\nonumber\\
\pi ^{24} (\det \sigma ) ^{12} \tet [\delta ]^8 \Psi _{10} ^2
&= &\left ( \prod _i \N _i ^4 \right )
\prod _{i<j} (a_i-a_j)^2 (b_i-b_j)^2
\eea
Here $\delta$ is an even spin structure. In the hyperelliptic
representation, it corresponds to a partition of the 6 branch points into two disjoint sets $\{a_1,a_2,a_3\}$ and $\{b_1,b_2,b_3\}$ of three branch points each. The first identity is a classic Thomae formula
\cite{mumford}, vol II, \S 8.
To establish the second identity, we make use of the following
bilinear $\tet$-constants, introduced in \cite{IV},
equation (2.38)
\be
{\cal M}_{\nu_i\nu_j}
=
\p_1\tet[\nu_i](0,\Omega)\,
\p_2\tet[\nu_j](0,\Omega)
-
\p_2\tet[\nu_i](0,\Omega)
\,
\p_1\tet[\nu_j](0,\Omega)
\ee
Solving for $\p_I\tet[\nu_i](0,\Omega)$ from
(\ref{proportionality}) and using the formula
(\ref{sigma}) for $\omega_I(z)$, we find
\be
(\det\,\sigma)\,{\cal M}_{\nu_i\nu_j}
=
4\pi^2\,\N_{\nu_i}\N_{\nu_j}(u_i-u_j)
\ee
Taking the products gives
\be
\label{M}
(\det\,\sigma)^4{\cal M}_{12}^2
{\cal M}_{23}^2
{\cal M}_{31}^2
{\cal M}_{45}^2
{\cal M}_{56}^2
{\cal M}_{64}^2
=
\prod_{i=1}^6\N_{\nu_i}^4
\prod_{i<j}(a_i-a_j)^2(b_i-b_j)^2
\ee
However, the ${\cal M}_{\nu_i\nu_j}^2$ have been determined completely explicitly in terms of $\tet$-constants in \cite{IV},
equation (4.9)
\be
{\cal M}_{\nu_i\nu_j}=\pi^4\tet[\delta]^2
\prod_{k\not=i,j}\tet[\nu_i+\nu_j+\nu_k]^2
\ee 
so that
\be
{\cal M}_{12}^2{\cal M}_{23}^2{\cal M}_{31}^2
=
{\cal M}_{45}^2{\cal M}_{56}^2{\cal M}_{64}^2
=
\pi^{12}\tet[\delta]^4
\Psi_{10}(\Omega).
\ee
Substituting this into (\ref{M}) gives the second identity
in (\ref{2identities}), and (\ref{2identities}) is now established. Taking the ratio of the two identities in
(\ref{2identities}), we find 
\be
\pi^3\,\Psi_{10}(\Omega)^{1/4} ~ \det\,\sigma
= \prod_{i=1}^6\N_{\nu_i}^{\half}
\ee
Comparing with
(\ref{dtau1}),
we obtain in this manner the following asymptotics for
$d \tau_1 \wedge d \tau_2$
\bea
\label{dtau2}
d \tau_1 \wedge d \tau_2
=
{1 \over 16 \pi ^3 } ~ t dt \wedge dp_2 ~ \omega _0 (p_1)^2
{ 1 \over \Psi _{10} ^{1/4}} \prod _\nu h_\nu (p_2)
+
\O(t^2).
\eea
The theorem is now an immediate consequence of
(\ref{Psi}), (\ref{dOmega}), and (\ref{dtau2}).
Q.E.D.

\subsection{Degeneration limit for the 3-loop bosonic string}

We recall that the genus $3$ bosonic string measure must satisfy the degeneration constraint (\ref{bosonic}).
Since the genus $2$ bosonic string measure is given by
$c_2\Psi_{10}^{-1}d^3\Omega$, Theorem 1 provides further evidence
that the genus $3$ bosonic string measure is given by
$c_3\Psi_9^{-1}(\Omega^{(3)})d^6\Omega^{(3)}$.
In fact, Theorem 1 also dictates what the coefficient of proportionality between the genera $2$ and $3$ must be
\be
c_3={c_2 \over (2\pi)^6}= { 1 \over 2^6 \pi ^{18}}
\ee 

As another check, we consider
the separating degeneration limit of the tachyon amplitude,
which is given by the following integral,
where $E(z,w)$ is the prime form,
\bea
\int \left | {dt \over t^2} \right |^2 \prod _{i<j} \left |
E(z_i,z_j) \right | ^{2 k_i \cdot k_j}
\eea
The behavior of the prime form when $z_i \in \Sigma^{(2)}$ and
$z_j \in \Sigma^{(1)}$ is given by
\bea
E(z_i,z_j) \to t^{- \half} E(z_i,p_2) E(p_1,z_j)
\eea
If the sum of the momenta on $\Sigma^{(2)}$ is $k$,
\bea
k= \sum _{i, ~z_i\in \Sigma^{(2)}} k_i =
- \sum _{j, ~z_j\in \Sigma^{(1)}} k_j
\eea
then we have the following $t$-dependence,
\bea
\int  \left | {dt \over t^2} \right |^2 |t|^{p^2}
\sim {1 \over p^2 -2}
\eea
which is the expected tachyon pole, with the correct value of the mass squared.

\newpage

\section{The genus 3 superstring measure as a degeneration problem}
\setcounter{equation}{0}

Using the formula for the genus $2$ superstring measure 
found in \cite{I} and the degeneration formulas of Theorem 1, 
we can formulate now more concretely the constraints on the 
genus $3$ superstring measure, in the degeneration limit where 
the worldsheet $\Sigma$ separates into a torus $\Sigma^{(1)}$ 
and a surface $\Sigma^{(2)}$ of genus $2$. Let $\Delta$ be an 
even genus $3$ spin structure. If, in this degeneration,
$\Delta$ factorizes into two odd spin structures, the leading contribution 
of order $t^{-2}$ to $d\mu[\Delta](\Omega^{(3)})$
vanishes, and we need not consider this case.
Henceforth, we assume  that $\Delta$ factorizes into
two even spin structures, and denote by $\delta_1$ and  
by $\delta_2\equiv\delta$ the even spin structures respectively 
on the torus and on the genus $2$ surface $\Sigma^{(2)}$.

\medskip

Let the genus $3$ superstring measure be expressed under the form
(\ref{Ansatz}),
for some density $\Xi_6[\Delta](\Omega^{(3)})$ yet
to be determined.
Recall that
in genus $h=2$, the superstring measure $d\mu[\delta](\Omega^{(2)})$ was shown to be given by \cite{I,IV}
\be
d\mu[\delta](\Omega^{(2)})
=
{\tet[\delta] ^4 (0,\Omega^{(2)})\,\Xi_6[\delta](\Omega^{(2)})
\over
16\pi^6\Psi_{10}(\Omega^{(2)})}\prod_{1\leq I\leq J\leq 2}d\Omega_{IJ}^{(2)}
\ee
The main expression $\Xi_6[\delta](\Omega^{(2)})$ is given in
\cite{I}, equation (7.1). We shall discuss it further in the next section. The degeneration constraint (\ref{superdeg}), Theorem 1,
and the degeneration formulas (\ref{thetaconst}) for
$\tet$-constants imply then that $\Xi_6[\Delta](\Omega^{(3)})$ must satisfy the following limit
\be
\label{condition1}
\lim_{t\to 0}
\Xi_6[\Delta](\Omega^{(3)})
=
\eta(\Omega^{(1)})^{12}
\
\Xi_6[\delta](\Omega^{(2)}).
\ee
This is the condition ({\it iii}) formulated in the Introduction.

\medskip

We discuss next the issue of modular invariance for 
$d\mu[\Delta](\Omega^{(3)})$. 
The full integrand in the amplitude (\ref{fullintegral})
must be invariant under $Sp(6,{\bf Z})$. Under the modular
transformations (\ref{modulartransformation}), we have
\bea
\label{modular1}
\det\, \Im \,\tilde\Omega^{(3)}&=& |\det\,(C\Omega^{(3)}+D)|^{-2}\
\det\, \Im\,\Omega^{(3)}
\nonumber\\
\prod_{I\leq J}
d\tilde\Omega_{IJ}^{(3)}
&=&
\det\,(C\Omega^{(3)}+D)^{-4}\
\prod_{I\leq J}d\Omega_{IJ}^{(3)}
\eea
At first sight, in genus $3$, we have difficulties due to the fact that 
the expression $\Psi_9(\Omega^{(3)})$
is defined only through its square, $\Psi_{18}(\Omega^{(3)})$.
However, the ambiguity in taking square roots here should not be 
relevant in string theory: for the superstring, $\Psi_9$ and its
conjugate appear in each chiral sector. This is also the case for the 
heterotic string, since we have seen that $\Psi_9$ appears
in the chiral measure for the bosonic string in the critical dimension, 
and this is unaffected by compactification. 
Thus the sign ambiguity in $\Psi_9$ can be ignored.
In analogy with the genus 2 case, we shall impose then the
modular transformation law ({\it ii}) described in the Introduction 
on the unknown term $\Xi_6[\Delta](\Omega^{(3)})$
\footnote{A slightly less restrictive requirement is to allow
in ({\it ii}) an additional phase $\epsilon(M)$ depending only
on the modular transformation $M$, but not on the spin structure
$\Delta$. Such additional phases do not affect significantly
our subsequent construction of candidates for $\Xi_6[\Delta](\Omega^{(3)})$.}. This condition implies that the
superstring chiral measure $d\mu[\Delta](\Omega^{(3)})$ transforms covariantly under modular transformations without any
phase factor
\be
d\mu[\tilde\Delta](\tilde\Omega^{(3)})
=
\det\,(C\Omega^{(3)}+D)^{-5}
\
d\mu[\Delta](\Omega^{(3)})
\ee
so that a manifestly modular invariant GSO projection is given by
\be
\sum_{\Delta}d\mu[\Delta](\Omega^{(3)}).
\ee
This completes our discussion of the three conditions
({\it i}-{\it iii}) formulated in the Introduction
for the modular covariant form $\Xi_6[\Delta](\Omega^{(3)})$. 

\newpage

\section{Ans\"atze for the superstring chiral measure}
\setcounter{equation}{0}

The goal of this section is to construct modular covariant
forms in genus $3$ satisfying the constraints ({\it ii})
and ({\it iii}). The starting point is the expression
$\Xi_6[\delta](\Omega^{(2)})$ in genus $2$. Our strategy is to
find and analyze analogous expressions in genus $3$. Although there are several natural analogues, it will turn out that the
degeneration condition ({\it iii}) is quite rigid,
and singles out a very small set of candidates.

\subsection{The form $\Xi_6[\delta](\Omega^{(2)})$ in genus $2$}

We begin by recalling the form $\Xi_6[\delta](\Omega^{(2)})$ in genus $2$. 
It was derived directly from the gauge-fixed genus $2$ superstring measure, 
and its original expression was heavily dependent on the fact that the 
worldsheet had genus $2$ (see \cite{I}, eq. (7.1)). More recently, two 
different expressions were found for $\Xi_6[\delta](\Omega^{(2)})$ 
which do admit generalizations to higher genus \cite{dp04}. 
To describe them, recall that a triplet $\{\delta_1,\delta_2,\delta_3\}$ 
is said to be {\sl asyzygous} if $e(\delta_1,\delta_2,\delta_3)=-1$ 
(respectively {\sl syzygous} when +1),
using the usual definitions of the signatures on pairs and triples of spin 
structures,
\bea
\label{triple}
\<\delta_i|\delta_j\> & \equiv & \exp (4\pi i(\delta_i'\delta_j''-\delta_i''\delta_j'))
\no \\
e(\delta_1, \delta_2, \delta_3) & \equiv & 
\<\delta_1|\delta_2\>\,\<\delta_2|\delta_3\>\,\<\delta_3|\delta_1\>
\eea
More generally,
we define as in \cite{dp04} an $N$-tuple of spin structures
of be {\it totally asyzygous}, 
if any triplet of distinct spin structures in the $N$-tuple
is asyzygous
\bea
&&
\{\delta_1,\cdots,\delta_N\} \ {\rm totally\ asyzygous}
\\ && \hskip .5in 
\Leftrightarrow
\
\{\delta_i,\delta_j,\delta_k\}
\ {\rm asyzygous},
\ \ {\rm for \ all} \ i,j,k \ {\rm pairwise\ distinct}.
\no
\eea
The notion of totally asyzygous $N$-tuple is modular invariant, since the cyclic product in (\ref{triple})
of relative signatures for a triple
of spin structures is. 

\medskip

Returning now to $\Xi_6[\delta](\Omega^{(2)})$, the first alternate expression
involves $4$-th powers of $\tet$-constants (as does the original expression in \cite{I}, eq. (7.1)) and is given by
\be
\label{Xi1}
\Xi_6[\delta](\Omega^{(2)})
=
-{1\over 2}\sum_{ { \{\delta,\delta_1,\delta_2,\delta_3\} \atop  {\rm tot.asyz.}} }
\
\prod_{i=1}^3\<\delta|\delta_i\>
\
\tet[\delta_i](0,\Omega^{(2)})^4
\ee
The notation indicates that, for given spin structure $\delta$, 
the summation runs over all triples $\{\delta_1,\delta_2,\delta_3\}$
such that $\{\delta,\delta_1,\delta_2,\delta_3\}$ forms
a totally asyzygous quartet. The second alternate expression
for $\Xi_6[\delta](\Omega^{(2)})$ involves only squares of $\tet$-constants, 
but it requires summation over certain sextets
of even spin structures. To identify which sextets, we define a
sextet $\{\delta_1,\cdots,\delta_6\}$ of spin structures in genus
$2$ to be {\it admissible} if it can be decomposed into three pairs
\be
\{\delta_1,\cdots,\delta_6\}
=
\{\delta_{i_1},\delta_{i_2}\}
\cup
\{\delta_{i_3},\delta_{i_4}\}
\cup
\{\delta_{i_5},\delta_{i_6}\},
\ee
with the union of any two pairs of the decomposition forming
a totally asyzygous quartet. For a given spin structure $\delta$,
we define the sextet $\{\delta_i\}$ to be {\it $\delta$-admissible} 
if it is admissible and it does not contain $\delta$. 
With this definition, the second alternative expression
for $\Xi_6[\delta](\Omega^{(2)})$ is given by
\be
\label{Xi2}
\Xi_6[\delta](\Omega^{(2)})
=
{1\over 2}\sum_{\{\delta_i\}
\,
\delta-adm.}
\
\e(\delta;\{\delta_i\})
\prod_{i=1}^6\tet[\delta_i](0,\Omega^{(2)})^2
\ee
Here the signs $\epsilon(\delta;\{\delta_i\}$ are uniquely related by modular transformations
\be
\label{phase2}
\epsilon(M\delta;\{M\delta_i\})
\
\prod_{i=1}^6\epsilon^2(\delta_i,M)
=
\epsilon^4(\delta,M)
\
\epsilon
(\delta;\{\delta_i\}),
\qquad
M\in Sp(4,{\bf Z}),
\ee
where $\epsilon^4(\delta,M)$ is the same factor occurring in the 
transformation law for $\tet^4[\delta]$. An explicit expression for the 
signs $\epsilon$ was given in \cite{dp04}.

\subsection{Ans\"atze in genus 3}

The preceding formulas for $\Xi_6[\delta](\Omega)$ in genus 2 suggest several natural extensions to genus~$3$. We discuss them below. The main issue is whether they can satisfy the
desired conditions ({\it i}), ({\it ii}), and ({\it iii})
listed in the Introduction, which are required for any viable
Ansatz for the genus $3$ superstring chiral measure.

\subsubsection{Ansatz in terms of asyzygous
quartets of spin structures}

The first alternate expression (\ref{Xi1}) for $\Xi_6[\delta](\Omega^{(2)})$ 
clearly makes sense for arbitrary genus, and in particular for genus $3$. 
Thus we are dealing here with an Ansatz  for $\Xi_6[\Delta](\Omega^{(3)})$ involving summations over spin structures $\{\Delta_1,\Delta_2,\Delta_3\}$
which together with the given spin structure $\Delta$,
form a totally asyzygous quartet. The modular covariant form which it defines
has been studied in \cite{dp04}, where it was denoted by
$\Xi_6^\#[\Delta](\Omega^{(3)})$. However, its degeneration limits, as 
determined in \cite{dp04}, Theorem 5, do not satisfy the degeneration constraint (\ref{condition1}) for the genus $3$ superstring measure.
Thus this Ansatz in terms of asyzygous quartets of spin structures 
must be dropped from contention. 

\subsubsection{Ans\"atze in terms of admissible sextets of spin structures}

We turn then to several Ans\"atze which can be viewed as generalizations to genus 3 of the expression (\ref{Xi2})
for $\Xi_6[\delta](\Omega)$ in terms of $\delta$-admissible sextets. First, in complete analogy with the genus case,
we define a sextet $\{\Delta_1,\cdots,\Delta_6\}$ to
be {\it admissible} if it can be decomposed as
\be
\label{admissible}
\{\Delta_{i_1},\Delta_{i_2}\}
\cup
\{\Delta_{j_1},\Delta_{j_2}\}
\cup
\{\Delta_{k_1},\Delta_{k_2}\}
\ee
with any two pairs constituting a totally asyzygous quartet.
Given spin structure $\Delta$, a sextet is said to be $\Delta$-{\it admissible} if it is admissible, and it does not
contain $\Delta$.

\medskip
Despite the similarity in the definitions, there is in practice a
fundamental difference between admissible sextets of spin 
structures in genus 2 and in genus 3: if 
\bea
s=
\{\delta_{i_1},\delta_{i_2}\}
\cup
\{\delta_{j_1},\delta_{j_2}\}
\cup
\{\delta_{k_1},\delta_{k_2}\}
\eea
is an admissible sextet in genus 2, then the triplets 
$\{ \delta _{i_\alpha}, \delta _{j_\beta}, \delta _{k_\gamma} \}$, with $\alpha,
\beta, \gamma =1,2$, are automatically syzygous.
Furthermore, if the sextet is $\delta$-admissible,
then the following triplet 
signatures are automatically determined, 
\bea
\label{genus2triplets}
e(\delta , e_{i _1}, e_{i_2}) = e(\delta , e_{j _1}, e_{j_2}) =
e(\delta , e_{k_1}, e_{k_2})=+1
\eea
This may easily be inferred by inspection of Table 4 in \cite{dp04}.

\medskip

This is no longer true for genus 3: in an admissible
sextet $\{\Delta_1,\cdots,\Delta_6\}$, the triplets 
$\{ \Delta _{i_\alpha}, \Delta _{j_\beta}, \Delta _{k_\gamma} \}$
need not all be syzygous (or all asyzygous). 
Thus the admissible sextets in genus 3 fall into 2 categories: 

\medskip 

(1) All triplets  $\{ \Delta _{i_\alpha}, \Delta _{j_\beta}, \Delta _{k_\gamma} \}$
are asyzygous, so that the whole sextet 
$\{\Delta_1,\cdots,\Delta_6\}$ is {\it totally asyzygous};

\medskip

(2) At least one triplet 
$\{ \Delta _{i_\alpha}, \Delta _{j_\beta}, \Delta _{k_\gamma} \}$
is syzygous. 
In this case, the relations  (\ref{genus2triplets}) also
do not follow
from $\Delta$-admissibility. In particular, one
has a classification depending on the following signs
\bea
\rho _1 & = & e(\Delta, \Delta _{j_1}, \Delta _{k_1})
\no \\
\rho _2 & = & e(\Delta, \Delta _{k_1}, \Delta _{i_1})
\no \\
\rho _3 & = & e(\Delta, \Delta _{i_1}, \Delta _{j_1})
\eea
The four resulting cases, namely $[+++]$, $[++-]$, $[+--]$ and $[---]$
are non-empty and the modular group acts within each case,
though not necessarily transitively so.
For convenience, we refer to all these cases as cases
of {\it partially asyzygous} sextets.

\newpage
$\bullet$ {\it Ans\"atze in terms of totally asyzygous sextets}

\medskip

We shall examine two Ans\"atze, in terms of totally asyzygous sextets with sign assignments.
\bea
\label{AB}
&({\rm A})&\ \ \Xi_6[\Delta](\Omega^{(3)})
\sim
\sum_{\{\Delta_i\} {\rm tot. \, asyz.}
\atop
\Delta\notin\{\Delta_i\}}
\e(\Delta;\{\Delta_i\})\prod_{i=1}^6\tet[\Delta_i](0,\Omega^{(3)})^2
\nonumber\\
&({\rm B})&\ \ \Xi_6[\Delta](\Omega^{(3)})
\sim
\bigg (
\sum_{\{\Delta_i\},\{\Delta_i'\} {\rm tot.\, asyz.}
\atop
\Delta\notin\{\Delta_i\},\{\Delta_i'\}}
\e(\Delta;\{\Delta_i\},\{\Delta_i'\})\prod_{i=1}^6\tet[\Delta_i](0,\Omega^{(3)})^2
\tet[\Delta_i'](0,\Omega^{(3)})^2
\bigg )^{1/2}
\nonumber\\
\eea
A key issue in these Ans\"atze is whether sign assignments exist
which are consistent with modular transformations. 
The first Ansatz (A) is simpler, and it will turn out that it does
satisfy the degeneration constraint (\ref{condition1}) if a consistent 
assignment existed. However, this turns out not to be the case, 
which is why the second Ansatz (B) is needed.
This second Ansatz (B) turns out to be the only viable candidate
for $\Xi_6[\Delta](\Omega^{(3)})$ among all the ones examined in
the present paper. Its full treatment requires the rest of
the paper. We postpone it then to the next section \S 6,
and complete now the discussion of the remaining Ans\"atze,
which involve partially asyzygous sextets.

\bigskip
$\bullet$ {\it Ansatz in terms of partially asyzygous sextets}

\medskip

In these remaining cases, the natural Ans\"atze would be   
\bea
\Xi_6[\Delta](\Omega^{(3)})
\sim
\sum_{[\rho_1,\rho_2,\rho_3]}
\e(\Delta;\{\Delta_i\})\prod_{i=1}^6\tet[\Delta_i](0,\Omega^{(3)})^2
\eea
where the summation would be over all $\Delta$-admissible sextets
$\{\Delta_1,\cdots,\Delta_6\}$ with some fixed
sign assignment $[\rho_1,\rho_2,\rho_3]$,
with not all $\rho_i$ equal to $-1$.

The first task is to examine whether consistent phase assignments $\epsilon(\Delta;\{\Delta_i\})$ exist. One does this orbit by
orbit under the modular group which leaves $\Delta$ invariant,
and uses the usual modular sign factors in the transformations 
of $\tet ^2$. The results are as follows,
where we have numbered the genus 3 even spin structures as
in Appendix \S C of \cite{dp04},

\begin{enumerate}
\item For the cases $[---]$, $[+--]$ and $[++-]$,
all orbits produce inconsistent sign assignments and are ruled out;
\item For the case $[+++]$, one orbit with 1680 sextets (generated by 
sextet $\{ 2,4,5,6,33,35\}$) and one orbit with 3360 sextets (generated by
sextet $\{ 5,7,12,13,22,30\}$) both generate {\sl consistent sign assignments};
\item For the case $[+++]$, there is one remaining orbit (generated by 
sextet $\{ 2,4,5,9,27,32\}$) which produces an inconsistent sign assignment.
\end{enumerate}

A simple example showing the non-existence of consistent phases for one of these orbits of $\Delta$-admissible, partially asyzygous sextets is given in section \S 6.2.3 below.

\medskip

The actual sums in both cases of 2. above are non-vanishing.
In the limit where the surface degenerates to a genus 2 times 
genus 1 surface, both sums converge to the same limit as 
the form $\Xi _6 ^\# [\delta ](\Omega ^{(3)})$
of \cite{dp04}, which is inconsistent with the requirement
({\it iii}) in the Introduction.
Thus, even though the sign assignments are consistent, 
the limits are not and the cases are all ruled out. 

\medskip
Although this analysis rules out a construction of $\Xi_6[\Delta](\Omega^{(3)})$ in terms of partially asyzygous $\Delta$-admissible sextets, it is in principle still possible that an Ansatz for
$\Xi_6[\Delta](\Omega^{(3)})^2$ can be obtained in terms of
pairs of partially asyzygous $\Delta$-admissible sextets,
just as we outlined in the preceding case (B) of totally
asyzygous $\Delta$-admissible sextets.
However, there does not appear to be any clear way of
recapturing $\Xi_6[\delta](\Omega^{(2)})^2$ from the degeneration limits of sums over pairs of partially asyzygous sextets.

\subsubsection{Ansatz in terms of dozens of spin structures}  

Since the previous Ans\"atze have not produced viable candidates 
for $\Xi_6[\Delta](\Omega^{(3)})$ itself as a polynomial in 
$\tet^2$, we may ask whether polynomials in $\tet$ could work,
\be
\Xi_6[\Delta](\Omega^{(3)})
=
\sum_{\{\Delta_i\}} \epsilon(\Delta;\{\Delta_i\}_{1\leq i\leq 12})
\prod_{i=1}^{12}\tet[\Delta_i](0,\Omega^{(3)}),
\ee
where the summation runs over a suitable set of dozens $\{\Delta_1,\cdots,\Delta_{12}\}$ of spin structures.
Here the expression for $\Xi_6[\delta](\Omega)$ in genus 2
provides little guidance for choosing this set.
There are conceivably many possibilities. But, 
given the good degeneration limits of totally asyzygous sextets,
it is natural to consider products of pairs of totally asyzygous 
sextets,
\bea
\Xi _6 ' [\Delta ](\Omega ^{(3)})
=
\sum _ {\{ s_1,s_2 \} \in {{\cal Q} _{pq}}} \epsilon (\Delta, s_1,s_2)
\prod _{i=1} ^{12} \tet [\Delta _i] (0,\Omega ^{(3)})
\eea
where the sum is over pairs of totally asyzygous sextets,
\bea
s_1  & = & 
\{ \Delta _1 , \Delta _2 , \Delta _3 , \Delta _4 , \Delta _5 , \Delta _6 \}
\no \\
s_2  & = & 
\{ \Delta _7 , \Delta _8 , \Delta _9 , \Delta _{10} , \Delta _{11} , \Delta _{12} \}
\eea
and ${\cal Q}_{pq}$ denote the different orbits of $\Delta$-admissible 
pairs of totally asyzygous sextets under the modular subgroup 
leaving $\Delta$ invariant. The orbits ${\cal Q}_{pq}$ are described in detail in section \S 6.3.2 below.

\medskip

Clearly, this construction can make sense only for orbits ${\cal Q}_{pq}$
for which the phases $\epsilon (\Delta, s_1,s_2)^2 $ can be consistently
defined. But this problem was already solved (by computer)
in the treatment of the Ansatz (B) in terms of pairs of totally
asyzygous sextets (see section \S 6.3.2 and subsequent discussions). It was found that consistent phases exist 
for the orbits ${\cal Q}_{01}$, ${\cal Q}_{02}$, ${\cal Q}_{13}$, ${\cal Q}_{20}$ ${\cal Q}_{21}$, ${\cal Q}_{22}$, and ${\cal Q}_{23}$ 
but not for the orbits 
${\cal Q}_{11}$, ${\cal Q}_{12}$, and ${\cal Q}_{3}$. A computer calculation
shows, however, that in none of these orbits the sign $\epsilon (\Delta, s_1,s_2)$
can actually be consistently defined.
Thus this particular Ansatz for $\Xi_6[\Delta](\Omega^{(3)})$ is also ruled out.

\newpage

\section{The Ans\"atze in terms of totally asyzygous sextets}
\setcounter{equation}{0}

To determine the degeneration behavior of the genus $3$ 
candidates (A) and (B), we need to determine the degeneration
behavior of genus $3$ totally asyzygous sextets of even spin structures.

\subsection{Degenerations of totally asyzygous sextets}

The basic fact is the following:

\bigskip

\noindent
{\bf Lemma 1}. {\it Let $\{\Delta_i\}_{1\leq i\leq 6}$ be a sextet of genus 
$3$ even spin structures. Assume that it is totally asyzygous and that
each $\Delta_i$ degenerates into even spin structures in genera 2 and 1. 
Let $\delta_1,\cdots,\delta_6$ be the 6 genus $2$ even spin structures 
which arise in this manner. Then the sextet $\{\delta_1,\cdots,\delta_6\}$
is an admissible sextet of genus $2$ even spin structures in the
sense defined above, that is, it can be divided into three pairs,
the union of any two defines a totally asyzygous quartet.
Furthermore, we have}
\be
\prod_{i=1}^6
\tet[\Delta_i](0,\Omega^{(3)})^2
\to
2^4\eta(\Omega^{(1)})^{12}
\prod_{i=1}^6
\tet[\delta_i](0,\Omega^{(2)})^2
\ee 

\bigskip
\noindent
{\it Proof.} We recall that there exist no totally asyzygous quintets at genus 2 
(and thus no genus 2 totally asyzygous sextuplets etc). 
This can be seen by direct inspection of the tables of asyzygies
in genus $2$ provided in \cite{dp04}.
Let $\mu_1,\cdots,\mu_6$ be the genus $1$ spin structures arising
from the degeneration of $\Delta_1,\cdots,\Delta_6$. 
By assumption, they are even. We examine in turn all possible arrangements for
$\mu_1,\cdots,\mu_6$ :

\begin{itemize}
\item Assume that $\mu_1 , \cdots , \mu_6$ take at most 2 distinct values
amongst the 3 possible even spin structures at genus 1. Then, it 
follows that $e(\mu_i,\mu_j,\mu_k)=+1$ for any triplet of $\mu$'s arising in the sextet. For the genus 3 sextet to be totally
asyzygous, the genus 2 sextuplet $\delta _1, \cdots , \delta _6$
must be totally asyzygous, but this is impossible.

\item Assume that five of the six $\mu_1 , \cdots , \mu_6$ (say 
$\mu_1 , \cdots , \mu_5$ for definiteness) take at  most 2 distinct 
values amongst the 3 possible even spin structures at genus 1. 
Then $e(\mu_i,\mu_j,\mu_k)=+1$ for $1\leq i,j,k \leq 5$. 
For the genus 3 sextet to be totally asyzygous, 
the genus 2 quintet $\delta _1, \cdots , \delta _5$
must be totally asyzygous, but this is impossible.

\item The only remaining possibility is that amongst the six 
$\mu_1 , \cdots , \mu_6$, each of the 3 distinct genus 1 even 
spin structures (which we denote $\mu_2 , \mu_3 , \mu_4$
by slight abuse of notation) occurs precisely twice. 
\end{itemize}
Thus, up to permutations of the $\mu$'s, we have
\bea
\label{genus3split}
\Delta _1 = \left ( \matrix{\delta _1 \cr \mu _2 \cr } \right )
\hskip 1in 
\Delta _3 = \left ( \matrix{\delta _3 \cr \mu _3 \cr } \right )
\hskip 1in
\Delta _5 = \left ( \matrix{\delta _5 \cr \mu _4 \cr } \right )
\no \\
\Delta _2 = \left ( \matrix{\delta _2 \cr \mu _2 \cr } \right )
\hskip 1in 
\Delta _4 = \left ( \matrix{\delta _4 \cr \mu _3 \cr } \right )
\hskip 1in
\Delta _6 = \left ( \matrix{\delta _6 \cr \mu _4 \cr } \right )
\eea
It is clear that the quartets
$\{\delta _1 ~ \delta _2 ~ \delta _3 ~ \delta _4\}$,
$\{\delta _1 ~ \delta _2 ~ \delta _5 ~ \delta _6\}$,
$\{\delta _3 ~ \delta _4 ~ \delta _5 ~ \delta _6\}$,
are totally asyzygous, and that they are the only totally asyzygous quartets within $\{\delta_1,\cdots,\delta_6\}$.
This proves the first part of Lemma 1. The second part
follows immediately from the degeneration formulas for
$\tet$-constants and from the identity (\ref{genus1identity}). Q.E.D.

\subsection{Orbits of sextets}

Since the genus $2$ expression $\Xi_6[\delta](\Omega^{(2)})$ is built from genus $2$ admissible sextets, Lemma 1 shows that asyzygous sextets have the potential to produce a form $\Xi_6[\Delta](\Omega^{(3)})$ tending
to $\Xi_6[\delta](\Omega^{(2)})$ in the degeneration limit.
Fix an even spin structure $\Delta$.
In analogy with the genus $2$ case, we 
define a $\Delta$-admissible sextet of even spin structures to be a totally asyzygous sextet $\{\Delta_i\}$ not containing $\Delta$. We can restrict then the sextets 
entering the candidate for $\Xi_6[\Delta](\Omega^{(3)})$ 
to the $\Delta$-admissible ones.
This justifies the form given in
(\ref{AB}) for the Ans\"atze (A) and (B).

\medskip
We need to consider the degenerations of $\Delta$-admissible sextets 
$\{\Delta_i\}$.  We can assume that
\be
\label{factorization}
\Delta=\pmatrix{\delta\cr \mu}
\ee
with both lower genus spin structures $\mu_i$ and $\delta$
even, since otherwise $\Delta$ will not contribute to the leading asymptotics. Let $\{\delta_i\}$ be the sextet of genus $2$ spin structures obtained by factoring $\{\Delta_i\}$.
We can assume that they are all even, since otherwise $\{\Delta_i\}$ will again not contribute to the leading asymptotics. Now
Lemma 1 guarantees that the sextet $\{\delta_i\}$ is admissible
in the genus $2$ sense. However, the condition $\Delta\notin\{\Delta_i\}$ does not guarantee that $\delta\notin \{\delta_i\}$, i.e., the $\Delta$-admissibility of
the genus $3$ sextet $\{\Delta_i\}$ does not guarantee the $\delta$-admissibility of the genus $2$ sextet $\{\delta_i\}$. 
Thus we have to analyze the contributions of genus $2$ admissible
sextets which are not $\delta$-admissible. We also have to
determine the exact multiplicities with which $\delta$-admissible and $\delta$-not admissible sextets occur
in the degeneration of an $Sp(6,{\bf Z})$ orbit of $\Delta$-admissible sextets in genus $3$. The first issue is addressed by Lemma 2 below. The second issue will be addressed by
a computer listing of all possibilities. The results will be described in subsequent sections.
 
\subsubsection{Orbits of admissible sextets in genus $2$}

The list of admissible sextets in genus $2$ is provided in \cite{dp04}, Table 4. 
By a simple inspection of that list and the actions of modular transformations 
in the same table, we find that

\medskip

$\bullet$ There are no totally asyzygous quintets,
and a fortiori, no totally asyzygous sextets in genus $2$;

$\bullet$ In genus $2$, there are 15 admissible sextets.
The group $Sp(4,{\bf Z})$ acts transitively on the set
of admissible sextets;

$\bullet$ Given a genus $2$ even spin structure $\delta$, there are always exactly $6$ sextets which are admissible, and $9$ which are not.
We denote these sets of sextets by $s[\delta]$ and $s^c[\delta]$
respectively
\bea
\label{S}
s[\delta]&=&\bigg\{\{\delta_i\} \ {\rm admissible\ sextet}\ ;\ 
\delta\notin \{\delta_i\}\bigg\}
\nonumber\\
s^c[\delta]&=&\bigg\{\{\delta_i\} \ {\rm  admissible\ sextet}\ ;
\ \delta\in \{\delta_i\}\bigg\};
\eea

$\bullet$ Let $Sp[\delta](4,{\bf Z})$ be the subgroup of 
$Sp(4,{\bf Z})$ fixing $\delta$. Then $Sp[\delta](4,{\bf Z})$
acts transitively on both $s[\delta]$ and $s^c[\delta]$.
In particular, if the phases $\epsilon(\delta;\{\delta_i\})$
satisfy the transformation (\ref{phase}), then all the phases
in each orbit $s[\delta]$ or $s^c[\delta]$ are uniquely 
determined by the phase of any single element inside 
$s[\delta]$ and $s^c[\delta]$.

\bigskip

\noindent
{\bf Lemma 2.} {\it Let $\delta$ be a fixed genus $2$ even spin structure. 
Assume that the phases $\epsilon(\delta;\{\delta_i\})$ satisfy the condition (\ref{phase2}) for all $M\in Sp[\delta](4,{\bf Z})$. Then we have
\bea
\label{SandSc}
\sum_{\{\delta_i\}\in s[\delta]}
\epsilon(\delta;\{\delta_i\})
\prod_{i=1}^6\tet[\delta_i]^2
&=&
\pm\,2\,\Xi_6[\delta](\Omega^{(2)})
\\
\sum_{\{\delta_i\}\in s^c[\delta]}
\epsilon(\delta;\{\delta_i\})
\prod_{i=1}^6\tet[\delta_i]^2
&=&
0.
\eea
The $\pm$ sign in the first identity is a consequence of the fact that the
phases $\epsilon(\delta;\{\delta_i\})$ in each orbit $s[\delta]$ or $s^c[\delta]$ 
are determined only up to a global sign.}

\bigskip
\noindent
{\it Proof.} The first identity in (\ref{SandSc}) is just a reformulation of (\ref{Xi2}), and was proved in \cite{dp04}.
To establish the second identity, we go to the hyperelliptic representation.

\medskip

Let $s^2=\prod_{i=1}^6(x-p_i)$ be a hyperelliptic representation
for the surface $\Sigma^{(2)}$
\footnote{The branch points $p_i$ here should not be confused with the punctures $p_1$ and $p_2$ in the degeneration construction of \S 2. The notation $p_i$ for the branch points
is in accord with \cite{dp04}, which is used heavily in the proof of Lemma 2.}. As before, we identify the spin structure $\delta$
with a partition of the 6 branch points into two sets of 3 branch points each, say
$\delta \sim  \{ a_1,a_2,a_3 \} \cup \{ b_1,b_2,b_3\}$.
The Thomae formula (for genus 2)  takes the following form,
\bea
\tet [\delta ] ^2 = \epsilon C
x_{a_1 a_2}  x_{a_2 a_3}  x_{a_3 a_1}
x_{b_1 b_2}  x_{b_2 b_3}  x_{b_3 b_1}
\hskip .5in 
x_{p_i p_j} = \sqrt{p_i-p_j}
\eea
Here, $\epsilon^4=1$, and $C$ is $\delta$-independent.
Actually, we need the explicit correspondence only for the 
sextets themselves. Given the normalization of a single
sextet, the correspondences for all others may be derived 
using the action of modular transformations on both sides.
We  fix the expression for one sextet, say $(125690)$, to be $C^6$,
and determine the 
hyperelliptic expressions for the others by modular transformations,
\bea
\label{deft}
t_1 \equiv + (125690) & = &  + (p_1-p_6) (p_2-p_4) (p_3-p_5) ~ C^6 V(p_i)
\no \\
t_2 \equiv + (137890) & = &  - (p_1-p_3) (p_2-p_5) (p_4-p_6) ~ C^6 V(p_i)
\no \\
t_3 \equiv + (145678) & = &  - (p_1-p_4) (p_2-p_3) (p_5-p_6) ~ C^6 V(p_i)
\no \\
t_4 \equiv + (124580) & = &  - (p_1-p_4) (p_2-p_6) (p_3-p_5) ~ C^6 V(p_i)
\no \\
t_5 \equiv + (134670) & = &  + (p_1-p_5) (p_2-p_3) (p_4-p_6) ~ C^6  V(p_i)
\no \\
t_6 \equiv + (123689) & = &  - (p_1-p_6) (p_2-p_5) (p_3-p_4) ~ C^6  V(p_i)
\no \\
t_7 \equiv - (134589) & = &  + (p_1-p_4) (p_2-p_5) (p_3-p_6) ~ C^6  V(p_i)
\no \\
t_8 \equiv - (124679) & = &  - (p_1-p_6) (p_2-p_3) (p_4-p_5) ~ C^6  V(p_i)
\no \\
t_9 \equiv - (123570) & = &  + (p_1-p_2) (p_4-p_6) (p_3-p_5) ~ C^6  V(p_i)
\no \\ 
t_{10} \equiv + (235678) & = & + (p_1-p_2)(p_3-p_4)(p_5-p_6) ~ C^6 V(p_i)
\no \\
t_{11} \equiv + (247890) & = & - (p_1-p_3)(p_2-p_6)(p_4-p_5) ~ C^6 V(p_i)
\no \\
t_{12} \equiv - (234579) & = & + (p_1-p_2)(p_3-p_6)(p_4-p_5) ~ C^6 V(p_i)
\no \\
t_{13} \equiv + (234680) & = & - (p_1-p_5)(p_2-p_6)(p_3-p_4) ~ C^6 V(p_i)
\no \\
t_{14} \equiv + (345690) & = & + (p_1-p_5)(p_2-p_4)(p_3-p_6) ~ C^6 V(p_i)
\no \\
t_{15} \equiv + (567890) & = & - (p_1-p_3)(p_2-p_4)(p_5-p_6) ~ C^6 V(p_i)
\eea
The omnipresent factor $V$ is the  Vandermonde polynomial
\bea
V (p_i) \equiv \prod _{1 \leq i < j \leq 6} x_{ij} ^2 
= \prod _{1 \leq i < j \leq 6} (p_i -p_j)
\eea
Under a permutation of the branch points, $V(p_i)$ is multiplied by 
the signature  of this permutation.
The following  modular transformations were used to 
establish these signs,
\bea
\Sigma (t_1) = + t_2 \hskip .5in &  
M_3 (t_4) = + t_8   & \hskip .5in 
S(t_{14}) = + t_{11} 
\no \\
T(t_1) = + t_3 \hskip .5in &  
S(t_8) = + t_7   & \hskip .5in 
M_1 (t_8) = - t_{12}
\no \\
S(t_3) = + t_6 \hskip .5in &  
T(t_7) = + t_9  & \hskip .5in 
M_3(t_{12}) = + t_{13}
\no \\
T(t_6) = + t_5  \hskip .5in &  
M_1 (t_3) = - t_{10}  & \hskip .5in 
S(t_{13}) = + t_{15}
\no \\
\Sigma (t_5) = + t_4 \hskip .5in &  
T(t_{10}) = + t_{14}  & 
\eea
Taking into account the behavior of $V$ under permutations, modular
invariance determines the relative signs in the sums over $s[\delta]$ and $s^c[\delta]$. Working this out for one of the spin structures, say $\delta = \delta_1$
gives the following explicit formulas,
\bea
\sum_{\{\delta_i\}\in s[\delta_1]}
\epsilon(\delta_1;\{\delta_i\}) \prod_{i=1}^6\tet[\delta_i]^2
& = & 2t_{12} + 2t_{13} + 2 t_{15} =   - 2t_{10} - 2t_{11} - 2t_{14}
\no \\ 
& = & 2 \Xi _6 [\delta _1]
\no \\
\sum_{\{\delta_i\}\in s^c[\delta_1 ]}
\epsilon(\delta_1;\{\delta_i\}) \prod_{i=1}^6\tet[\delta_i]^2
& = & \sum _{i=1} ^ 9 t_i =0
\eea
The modular covariance properties then yield these results for all
spin structures $\delta$ and thus completes the proof of the theorem.
Q.E.D.

\subsubsection{Orbits of admissible sextets in genus $3$ (totally asyzygous sextets)}

We list here a number of results on the modular transformations
of asyzygous multiplets $\{\Delta_i\}$ in genus $3$, all of which 
have been proven by computer calculations.

\medskip

$\bullet$ The sets of all asyzygous quartets, quintets and sextets transform
transitively under the full modular group acting on characteristics;

\medskip

$\bullet$ There are 5040 totally asyzygous quartets, 2016 totally asyzygous 
quintets, $336$ totally asyzygous sextets,  and no totally asyzygous septets;

\medskip

$\bullet$ The set of all asyzygous sextets that do not contain a given 
spin structure $\Delta$ transforms transitively
under the modular subgroup $Sp[\Delta](6,{\bf Z})$
leaving $\Delta$ invariant. In analogy with the genus $2$ case,
we denote by $S[\Delta]$ the set of asyzygous sextets not
containing a spin structure $\Delta$. For any $\Delta$,
$S[\Delta]$ consists of $280$ elements;

\medskip

$\bullet$ Upon factorization, each sextet $\{\Delta_i\}$ of genus 3 spin structures produces a sextet $\{\delta_i\}$ of genus 2
spin structures. Consider the set of the 336 sextets $\{\delta_i\}$ of genus
2 spin structures which are obtained from factorization from
the set of all $336$ asyzygous sextets in genus 3.
Then the set of such $\{\delta_i\}$
can be divided into 246 sextets which contain at least some odd spin structure, together with 6 copies
of all 15 genus 2 admissible sextets;

\medskip 

$\bullet$ 
Similarly, let $\Delta$ factorize into a genus $1$ and a genus $2$ 
spin structure $\delta$ as in (\ref{factorization}),
and consider the set of all genus $2$ sextets $\{\delta_i\}$ 
arising from factorization of the 280 $\Delta$-admissible
genus 3 sextets in $S[\Delta]$. Then the set of such $\{\delta_i\}$
can be divided into 208 sextets which contain at least some odd
spin structures, together with 6 copies of $s[\delta]$ and 
4 copies of $s^c[\delta]$. 

\medskip

We can now consider the first Ansatz in (\ref{AB}) for $\Xi_6[\Delta](\Omega^{(3)})$, where the summation is over the
set $S[\Delta]$ of $\Delta$-admissible sextets. For 
$\Xi_6[\Delta](\Omega^{(3)})$ to transform as in ({\it ii}), we impose the analogous condition to (\ref{phase2}) in genus $3$ 
\be
\label{phase3}
\epsilon(M\Delta;\{M\Delta_i\})
\
\prod_{i=1}^6\epsilon^2(\Delta_i,M)
=
\epsilon^4(\Delta,M)
\
\epsilon
(\delta;\{\Delta_i\}),
\qquad
M\in Sp(6,{\bf Z}),
\ee
Restricted to $M\in Sp[\Delta](6,{\bf Z})$, this implies that all the phases $\epsilon(M\Delta;\{M\Delta_i\})$ in the first Ans\"atz uniquely determine one another. Assuming the 
existence of such a consistent assignment of phases, the expression in the first Ans\"atz is then uniquely determined up
to a global $\pm$ sign. The $Sp[\Delta](6,{\bf Z})$ consistency of phases implies the $Sp[\delta](4,{\bf Z})$ consistency of phases. Thus Lemmas 1 and 2 apply. Together with the numerology for the degeneration of the orbit $S[\Delta]$ found above,
we obtain
\be
\lim_{t\to 0}
\sum_{\{\Delta_i\}\in S[\Delta]}
\epsilon(\Delta;\{\Delta_i\})
\prod_{i=1}^6\tet[\Delta_i]^2(0,\Omega^{(3)})
=
6\cdot
2^4\,
\eta(\Omega^{(1)})^{12}\,
\Xi_6[\delta](\Omega^{(2)})
\ee
Here, we assume that all 6 copies of $s[\delta]$ obtained in factoring 
$S[\Delta]$ lead to contributions of the same sign. However, there is a 
more severe obstruction to the Ansatz (A):

\medskip

$\bullet$ There does not exist a phase assignment
$\epsilon(\Delta;\{\Delta_i\})$ satisfying the condition
(\ref{phase3}) and the sextets are totally asyzygous.
 This is in marked contrast with the genus $2$
case, where the phases $\epsilon(\delta;\{\delta_i\})$ satisfying (\ref{phase2}) do exist. A counterexample in genus $3$ is obtained by
considering the following $\Delta_1$-admissible sextet,\footnote{Throughout,
we shall use the nomenclature for genus 3 spin structures and 
modular transformations given in Appendix C of \cite{dp04}.}
\bea
s_1 = (\Delta _2, \Delta _8, \Delta _{14}, \Delta _{16}, \Delta _{25}, \Delta _{30})
\eea
and the action of the composite modular transformation $A_1B_4$.
From Table 6 of \cite{dp04}, it is clear that $A_1B_4$  leaves 
$\Delta_1, \Delta _3, \Delta _4, \Delta _6$ invariant
and maps $\Delta _2 \leftrightarrow \Delta _5$. 
Thus, the $\Delta$-admissible sextet $s_1$, as a whole, is invariant 
under $A_1B_4$. The sign factor is also easily computed, using
\bea
\epsilon ^2  (\Delta _i, A_1 B_4) 
= \epsilon ^2 (B_4 \Delta _i, A_1) \times \epsilon ^2 (\Delta _i, B_4)
= e^{4 \pi i (\Delta _i)_1 ' (\Delta _i)_2'}
\eea
and we find 
\bea
\epsilon ^2 (\Delta _i, A_1 B_4) = +1 \hskip .5in i=2,8,14,16,25 \hskip .5in 
\epsilon ^2 (\Delta _{30}, A_1 B_4) = -1
\eea
But then the sextet contribution changes sign under a transformation that 
leaves the sextet invariant, which means to no consistent sign can be defined.

\subsubsection{Orbits of admissible sextets in genus 3 (partially asyzygous sextets)}

A consistent phase assignment is also lacking in this case. A counterexample
in genus 3 is obtained by considering the following $\Delta_1$-admissible sextet,
\bea
s_2 = (\Delta _2, \Delta _6, \Delta _8, \Delta_{18}, \Delta_{29}, \Delta _{36})
\eea
The  modular transformation $A_6B_6A_6B_6$ leaves each of the spin
structures in $s_2$, and thus the entire sextet, invariant. The signs 
accompanying the transformation are easily computed, using
\bea
\epsilon ^2  (\Delta _i, A_6B_6A_6B_6)  = +1 & \qquad & i=2,8,18 
\no \\
\epsilon ^2  (\Delta _i, A_6B_6A_6B_6)  = -1 & \qquad & i=6,29,36 \hskip .5in 
\eea
But then the sextet contribution changes sign under a transformation that 
leaves the sextet invariant, which means to no consistent sign can be defined.

\subsection{Orbits of pairs of sextets}

In the preceding section, we have seen sums over $\Delta$-admissible 
sextets are not consistent with the
modular transformation (\ref{phase3}). Thus we cannot construct 
$\Xi_6[\Delta](\Omega^{(3)})$ directly by the Ansatz (A). In this section, we shall show that certain sums over {\it pairs of sextets} do admit consistent phase assignments, and that carefully chosen sums do lead to viable candidates for $\Xi_6[\Delta](\Omega^{(3)})^2$.

\subsubsection{Orbits of pairs of admissible sextets in genus 2}

Fix an external  genus $2$ even  spin structure $\delta$.
Our first task is to identify the orbits of pairs 
$\delta$-admissible sextets under $Sp[\delta](4,{\bf Z})$.
Clearly, for each integer $p$, the subset of pairs
$\{\delta_i\}$, $\{\delta_i'\}$ with $p$ common spin structures
is invariant under $Sp[\delta](4,{\bf Z})$. For $\delta$-admissible pairs of sextets, there is a finer partition
which does give precisely all the orbits under $Sp[\delta](4,{\bf Z})$:
\bea
{Q}_p^{0,0}[\delta]&=&
\big\{ (\{\delta_i\},\{\delta_i'\})\in s[\delta]\times s[\delta]; \ \#(\{\delta_i\}\cap\{\delta_i'\})=p\big\}
\nonumber\\
{Q}_p^{0,1}[\delta]&=&
\big\{ (\{\delta_i\},\{\delta_i'\})\in s[\delta]\times s^c[\delta]; \ \#(\{\delta_i\}\cap\{\delta_i'\})=p\big\}
\nonumber\\
{Q}_p^{1,0}[\delta]&=&
\big\{ (\{\delta_i\},\{\delta_i'\})\in s^c[\delta]\times s[\delta]; \ \#(\{\delta_i\}\cap\{\delta_i'\})=p\big\}
\nonumber\\
{Q}_p^{1,1}[\delta]&=&
\big\{ (\{\delta_i\},\{\delta_i'\})\in s^c[\delta]\times s^c[\delta]; \ \#(\{\delta_i\}\cap\{\delta_i'\})=p\big\}
\eea

By inspecting the table of admissible sextets in genus $2$, we find that only the values $p=3,4$ and $6$ produce non-empty
sets $Q_p^{a,b}[\delta]$. 
The sizes of the orbits $Q_p^{a,b}[\delta]$ are
given by
$\#\,Q_3^{0,0}[\delta]=12$,
$\#\,Q_4^{0,0}[\delta]=18$,
$\#\,Q_6^{0,0}[\delta]=6$,
$\#\, Q_3^{0,1}[\delta]=\#\,Q_3^{1,0}[\delta]=36$,
$\#\, Q_4^{0,1}[\delta]=\#\,Q_4^{1,0}[\delta]=18$,
$\#\,Q_3^{1,1}[\delta]=36$,
$\#\,Q_4^{1,1}[\delta]=36$,
$\#\,Q_6^{1,1}[\delta]=9$,
which does add up to $15^2=225$. In this counting, the pairs
of sextets have been viewed as ordered pairs. For later purposes,
it is preferrable to count unordered pairs, in which case the
sizes of the orbits $Q_p^{a,b}[\delta]$ become
\be
\matrix{Q_3^{0,0}=6\qquad &Q_3^{0,1}=36\qquad&Q_3^{1,1}=18\cr
Q_4^{0,0}=9\qquad&Q_4^{0,1}=18\qquad&Q_4^{1,1}=18\cr
Q_6^{0,0}=6\qquad&Q_6^{0,1}=0\qquad&Q_6^{1,1}=9\cr}
\ee

\medskip
To each orbit $Q_p^{a,b}[\delta]$, we can associate 
the following polynomial in $\tet$-constants 
\bea
F_p^{a,b}[\delta]=
\sum_{(\{\delta_i\},\{\delta_i'\})\in Q_p^{a,b}[\delta]}
\e_p^{a,b}(\delta;\{\delta_i\},\{\delta_i'\})
\prod_{i=1}^6\tet[\delta_i]^2
\prod_{i=1}^6\tet[\delta_i']^2
\eea
where the phases $\e(\delta;\{\delta_i\},\{\delta_i'\})$ are required to satisfy
\be
\label{phase2pairs}
\e(M\delta;\{M\delta_i\},\{M\delta_i'\})
\prod_{i=1}^6\e^2(\delta_i,M)
\prod_{i=1}^6\e^2(\delta_i',M)
=
\e(\delta;\{\delta_i\},\{\delta_i'\})
\ee
Since $Q_p^{a,b}[\delta]$ are orbits of $Sp[\delta](4,{\bf Z})$,
the phases $\e(\delta;\{\delta_i\},\{\delta_i'\})$ completely determine each other within $Q_p^{a,b}[\delta]$. We also find,
by computer inspection, that a consistent assignment of phases
$\e(\delta;\{\delta_i\},\{\delta_i'\})$ exist for each $Q_p^{a,b}[\delta]$. Thus the expressions $F_p^{a,b}[\delta]$
exist, and are uniquely determined by a single normalizing
sign. We shall define this normalizing sign below.

\medskip
Remarkably, the expressions $F_p^{a,b}[\delta]$ can be expressed 
very simply in terms of $\Xi_6[\delta](\Omega^{(2)})$ and two
other polynomials in $\tet$-constants, defined by
\bea
F [\delta_1] \equiv \sum _{i \in s [\delta_1 ]} t_i ^2
& \hskip .6in &
s[\delta_1] = \{ 10, 11, 12, 13, 14, 15 \}
\no \\
F^c [\delta_1] \equiv \sum _{i \in s^c [\delta_1]} t_i ^2
& \hskip .6in &
s^c [\delta _1]  = \{ 1,2,3,4,5,6,7,8,9 \}
\eea
Then we have

\bigskip
\noindent
{\bf Lemma 3.} {\it Let the normalizing signs for
$F_p^{a,b}[\delta]$ be defined by the equation
(\ref{normalizingsign}). Then the expressions $F_p^{a,b}[\delta]$
are given by}
\bea
\label{Fexp}
F^{0,0} _3 [\delta_1]  =   \Xi_6 [\delta _1]^2 -  F [\delta _1]/2
\qquad
& F^{0,1} _3 [\delta_1]  = - F^c [\delta _1] &
\qquad 
F^{1,1} _3 [\delta_1] =    F^c [\delta_1] /2
\no \\ 
F^{0,0} _4 [\delta_1]  =  - \Xi _6 [\delta _1] ^2
\hskip .9in
& F^{0,1} _4 [\delta_1]  = F^c [\delta _1] &
\qquad
F^{1,1} _4 [\delta_1]  =  - F^c [\delta_1]
\no \\
F^{0,0} _6 [\delta_1]  = F [\delta _1] 
\hskip 1.15in & & \qquad
F^{1,1} _6 [\delta_1]  =  + F^c [\delta _1] \qquad
\eea

\medskip
\noindent
{\it Proof.} 
The following relations were established earlier,
\bea
\label{rel1}
\Xi_6 [\delta_1] = t_{10} + t_{11} + t_{14}
= - t_{12} - t_{13} - t_{15}
\eea
\bea
\label{rel2}
0 = t_1 + t_2 + t_3 + t_4 + t_5 + t_6 + t_7 + t_8 + t_9
\eea
The relation (\ref{rel1}) was established as a step in the proof
of the alternative form (\ref{Xi2}) of $\Xi_6[\delta](\Omega^{(2)})$ in \cite{dp04}.
The relation (\ref{rel2}) is a reformulation of the second identity in Lemma 2.
Additional ``rearrangement" formulas are as follows, 
\bea
\label{rel3}
t_1 & = & t_2 + t_3 + t_5 + t_7  ~=~ t_{14} + t_{15}   
\no \\
t_2 & = & t_1 + t_3 + t_4 + t_8 ~=~ t_{11} + t_{15}
\no \\
t_3 & = & t_1 + t_2 + t_6 +t_9 ~=~ t_{10} + t_{15}      
\no \\
t_4 & = & t_2 + t_5 + t_6 +t_8 ~=~ t_{11} + t_{13}      
\no \\
t_5 & = & t_1 + t_4 +t_6 +t_7 ~=~  t_{13} + t_{14}      
\no \\
t_6 & = & t_3 + t_4 +t_5 +t_9 ~=~ t_{10} + t_{13}       
\no \\
t_7 & = & t_1 +t_5 + t_8 +t _9 ~=~ t_{12} + t_{14}  
\no \\
t_8 & = & t_2  + t_4  + t_7 + t_9 ~=~  t_{12} + t_{11}   
\no \\
t_9 & = & t_3 + t_6 +t_7 +t_8 ~=~  t_{12} + t_{10}       
\eea
and 
They follow directly from the hyperelliptic representation; the equivalences are under the relations (\ref{rel1}, \ref{rel2}).

\medskip
We define now the normalizing signs for $F_p^{a,b}[\delta]$ promised earlier.
Writing $e_p^{a,b}(\delta_1;t_i,t_j)=
e_p^{a,b}[\delta](i,j)$ for simplicity, they are given by
\bea
\label{normalizingsign}
\epsilon _3 ^{0,0} [\delta_1](10,11)    = +1 & \hskip .3in &
\epsilon _3 ^{0,1} [\delta_1](1,10)    = +1  \hskip .5in
\epsilon _3 ^{1,1} [\delta_1](1,2)  = +1 
\no \\
\epsilon _4 ^{0,0} [\delta_1](10,13)    = +1 & \hskip .3in &
\epsilon _4 ^{0,1} [\delta_1](1,14)    = +1  \hskip .5in 
\epsilon _4 ^{1,1} [\delta_1](1,4)      = +1 
\no \\
\epsilon _6 ^{0,0} [\delta_1](10,10)    = +1 & \hskip .3in & \hskip 1.81in
\epsilon _6 ^{1,1} [\delta_1](1,1)  = +1 
\no \\ && 
\eea
The resulting polynomials are then as follows,
\bea
F^{0,0} _3 [\delta_1](t) 
& = & 
+ t_{10} t_{11} + t_{10} t_{14} + t_{11} t_{14} 
+ t_{12} t_{13} + t_{12} t_{15} + t_{13} t_{15}
\no \\ 
F^{0,0} _4 [\delta_1](t) 
& = & 
+ (t_{10} + t_{11} + t_{14} )(t_{12} + t_{13} + t_{15} )
\no \\
F^{0,0} _6 [\delta_1](t) 
& = & 
+ t_{10} ^2 + t_{11}^2 + t_{12}^2 + t_{13}^2 + t_{14}^2 + t_{15}^2
\no \\
F^{0,1} _3 [\delta_1](t) 
& = & 
+ t_1 (t_{10} + t_{11} + t_{12} + t_{13})
+ t_2 (t_{10} + t_{12} + t_{13} + t_{14})
\no \\ &&
+ t_3 (t_{11} + t_{12} + t_{13} + t_{14})
+ t_4 (t_{10} + t_{12} + t_{14} + t_{15})
\no \\ &&
+ t_5 (t_{10} + t_{11} + t_{12} + t_{15})
+ t_6 (t_{11} + t_{12} + t_{14} + t_{15})
\no \\ && 
+ t_7 ( t_{10} + t_{11} + t_{13} + t_{15})
+ t_8 ( t_{10} + t_{13} + t_{14} + t_{15})
\no \\ &&
+ t_9 (t_{11} + t_{13} + t_{14} + t_{15})
\no \\
F^{0,1} _4 [\delta_1](t) 
& = & 
+t_1 (t_{14} + t_{15}) + t_2 (t_{11} + t_{15}) + t_3 (t_{10} + t_{15})
\no \\ &&
+ t_4 (t_{11} + t_{13})  + t_5 (t_{13} + t_{14})  + t_6 (t_{10} + t_{13})
\no \\ && 
+ t_7 (t_{12} + t_{14}) + t_8 (t_{11} + t_{12}) + t_9 (t_{10} + t_{12})
\no \\
F^{1,1} _3 [\delta_1](t) 
& = & 
+t_1 t_2 + t_1 t_3 + t_1 t_5 + t_1 t_7 + t_2 t_3 + t_2 t_4 + t_2 t_8
+ t_3 t_6 + t_3 t_9 
\no \\ && 
+ t_4 t_5 + t_4 t_6 + t_4 t_8 + t_5 t_6 + t_5 t_7
+ t_6 t_9 + t_7 t_8 + t_7 t_9 + t_8 t_9
\no \\
F^{1,1} _4 [\delta_1](t) 
& = & 
+ t_1 t_4 + t_1 t_6 + t_1 t_8 + t_1 t_9 + t_2 t_5 + t_2 t_6 + t_2 t_7 + t_2 t_9
+ t_3 t_4
\no \\ &&
+ t_3 t_5 + t_3 t_7 + t_3 t_8 + t_4 t_7 + t_4 t_9 + t_5 t_8 + t_5 t_9
+ t_6 t_7 + t_6 t_8
\no \\
F^{1,1} _6 [\delta_1](t) 
& = & 
+ t_1^2 + t_2 ^2 + t_3 ^2 + t_4 ^2 + t_5 ^2 + t_6 ^2 + t_7 ^2 + t_8 ^2 + t_9^2
\eea
The above expressions for $F$ can be recast in the following,
more systematic way,
\bea
F^{0,0} _p [\delta_1](t) 
& = & 
\sum _{{\#(i\cap j)=p, \atop i\leq j; i,j \in s[\delta_1] }}  t_i t_j
\hskip 1in p=3,4,6
\no \\
F^{0,1} _p [\delta_1](t) 
& = & 
\sum _{{\#(i\cap j)=p, \atop i \in s[\delta_1], j \in s^c[\delta_1] }}  t_i t_j
\hskip 1in p=3,4
\no \\
F^{1,1} _p [\delta_1](t) 
& = & 
\sum _{{\#(i\cap j)=p, \atop i\leq j, i,j \in s^c[\delta_1]}}  t_i t_j
\hskip 1in p=3,4, 6
\eea
Using the relations (\ref{rel1}), (\ref{rel2}), and (\ref{rel3}),  we can reduce 
these  expressions to the linear combinations of the 3 standard forms
$\Xi_6[\delta_1]^2$, $F[\delta_1]$, and $F^c [\delta _1]$ given in Lemma 3. Q.E.D.

\subsubsection{Orbits of pairs of admissible sextets in genus 3}

We consider next the same issue of orbits and consistency of phase assignments for pairs of admissible sextets in genus $3$.
The following can be found by computer listings:

\medskip

$\bullet$
The set of all pairs of asyzygous sextets may be decomposed 
into 7 mutually exclusive sets, according to whether the two sextets
in the pair have 0, 1, 2, 3, 4, 5, or 6 spin structures in common.
Each of these sets of pairs transforms transitively under the group of all
modular transformations $Sp(6,{\bf Z})$.
The number of pairs in each category is listed in the second column of 
the table below. 

\medskip

Also, we shall need the number of pairs of sextets,
such that neither sextet in the pair contains a given spin structure $\Delta_1$.
The numbers of such pairs in each category is listed in the third
column of the table below. Under modular subgroup that preserves 
$\Delta_1$,  the sets with 0, 1, and 2 spin structures in common  are 
NOT transitive. The table below lists in the fourth column the sizes
of the orbits of the modular subgroup $Sp[\Delta_1](6,{\bf Z})$.

\begin{table}[htdp]
\begin{center}
\begin{tabular}{|c||c|c|c|c|}
\hline
\# $\cap$ & \# pairs & \# pairs  $ \not\supset \Delta_1$ & Orbits  & 
Reference pair \\ 
\hline \hline
0 & 15120 & 10080 & 5400$^{(1)}$     
    & $\{ 2,15,17,19,22,32\}, ~ \{ 9,10,13,16,27,28 \} $\\ \hline 
 & & & 5400$^{(2)}$     
    &  $\{ 4,5,10,16,26,36\}, ~ \{ 7,11,14,20,27,33 \}$ \\ \hline \hline
1 & 30240 & 21000 & 840                    
    & $\{ 3,4,12,17,25,29\}, ~ \{10,20,22,25,28,35\}$ \\ \hline
 & & & 10080$^{(1)}$   
    &  $\{ 5,15,22,27,29,31\}, ~ \{ 4,11,18,19,24,31\}$ \\ \hline 
 & & & 10080$^{(2)}$  
    &  $\{ 2,5,17,19,28,30\}, ~ \{ 2,8,10,15,25,32\}$ \\ \hline \hline
2 & 7560 & 5460 & 1260                 
    &   $\{ 8,12,15,20,28,33 \}, ~ \{ 4,12,13,17,24,33 \}$ \\ \hline 
 & & & 1680                 
    &  $ \{ 4,5,10,16,26,36\}, ~ \{ 3,4,7,12,22,36 \}$   \\ \hline 
 & & & 2520                 
    &  $ \{ 5,14,16,20,25,35 \}, ~ \{ 6,16,24,25,30,36 \}$\\ \hline \hline
3 & 3360 & 2800 & 2800              
    &  $ \{ 1,4,10,17,27,33 \}, ~ \{ 4,10,11,17,25,26 \} $ \\ \hline \hline
4 & 0 & 0 & --              
    &  --  \\ \hline \hline
5 & 0 & 0 & --               
    &  --  \\ \hline \hline
3 & 336 & 280 & 280              
    &  any pair \\ \hline 
\end{tabular}
\end{center}
\label{table:1}
\caption{Numbers of pairs of asyzygous sextets and modular orbits excluding $\Delta_1$}
\end{table}%

$\bullet$ 
Consider the transformation law for sign assignments 
$\epsilon(\Delta;\{\Delta_i\},\{\Delta_i'\})$ for pairs of sextets in 
genus $3$ given by the analogue of (\ref{phase2pairs}),
\be
\label{phase3pairs}
\e(M\Delta;\{M\Delta_i\},\{M\Delta_i'\})
\prod_{i=1}^6\e^2(\Delta_i,M)
\prod_{i=1}^6\e^2(\Delta_i',M)
=
\e(\Delta;\{\Delta_i\},\{\Delta_i'\}),
\ee 
where $M$ is any element of $Sp(6,{\bf Z})$.
With computer calculations, using all the generators 
$S$, $M_{A_i}$, and $M_{B_i}$, $i=1, \cdots , 6$
of the full $Sp(6,{\bf Z})$, the following may be shown.
\begin{enumerate}
\item  A unique (up to a global sign) and consistent sign assignment
exists for all the orbits in the sets with 0, 2 and 6 spin structures in common, 
as well as for the orbit $10080^{(2)}$ in the set with 1 spin structure in common;
\item No consistent sign assigment exists for any of the  other orbits.
\end{enumerate}

\subsection{Branching rules for $Sp[\Delta](6,{\bf Z})$ orbits
into $Sp[\delta](4,{\bf Z})$ orbits}

In this section, we list the multiplicities of all the $Sp[\delta](4,{\bf Z})$ 
orbits which arise upon factorization of the orbits of $Sp[\Delta](6,{\bf Z})$. 
Recall that, in genus $3$, the invariant set of pairs of $\Delta$-admissible 
sextets with $p$ common spin structures can be decomposed further into
irreducible orbits. Let these orbits be denoted by ${\cal Q}_{pq}$, with $p$ indicating that the pairs of $\Delta$-admissible sextets
have $p$ common spin structures, and $q$ indicating 
which orbit is being considered for given $p$. 
In the table below,  $N_{pq}$ denotes the multiplicity of the genus 2 orbit in the decomposition of the orbit ${\cal Q}_{pq}$. Also, $\# (Q)$ denotes the cardinality of the genus 2 even spin structure orbit.

\begin{table}[htb]
\begin{center}
\begin{tabular}{|c|c||c|c|c|c|c|c|} \hline 
genus 2 orbit & $\#(Q)$  & $N_{01}$ & $N_{02}$ & $N_{21}$ & $N_{22}$ & $N_{23}$ & $N_{6}$
                \\ \hline \hline
              $Q^{0,0} _3$  & 6
            & 12
            & 0     
            & 0
            & 0 
            & 0
            & 0
 \\ \hline
              $Q^{0,1} _3$ & 36
            & 4
            & 4   
            & 0 
            & 0       
            & 0
            & 0 
 \\ \hline
              $Q^{1,1} _3$ & 18
            & 0 
            & 4        
            & 0 
            & 0     
            & 0
            & 0 
 \\ \hline
              $Q^{0,0} _4$   & 9
            & 0 
            & 0        
            & 0 
            & 4      
            & 8
            & 0 
 \\ \hline
              $Q^{0,1} _4$ & 18
            & 0 
            & 0        
            & 8 
            & 0       
            & 0
            & 0 
 \\ \hline
              $Q^{1,1} _4$  & 18
            & 0 
            & 0        
            & 2 
            & 2       
            & 2
            & 0 
 \\ \hline
              $Q^{0,0} _6$   &6
            & 0 
            & 6        
            & 6 
            & 3     
            & 0
            & 6 
 \\ \hline
              $Q^{0,1} _6$ & 0
            & 0 
            & 0        
            & 0 
            & 0       
            & 0
            & 0 
 \\ \hline
              $Q^{1,1} _6$  & 9
            & 2 
            & 0        
            & 2 
            & 0       
            & 2
            & 4 
 \\ \hline \hline
 Total number of pairs & & 234 & 252 & 234 & 90 & 126 & 72 
 \\ \hline \hline
\end{tabular}
\end{center}
\caption{Branching rules for genus 3 orbits into genus 2 orbits}
\label{table:18}
\end{table}

\bigskip

The computer analysis also shows that, in the above table,
all copies of any given orbit $Q_p^{a,b}[\delta]$ always occur with the sign $+$.
Thus there is no cancellation between the various copies
of any orbit $Q_p^{a,b}[\delta]$. (Of course, the global
sign in front of each $F_p^{a,b}[\delta]$ is a matter of
convention, depending on the choice of global sign for
the definition of $F_p^{a,b}[\delta]$. 

\medskip

To each $Sp[\Delta](6,{\bf Z})$ orbit ${\cal Q}_{pq}$, we can associate then a polynomial $P_{pq}$ in genus $2$ $\tet$-constants, defined
as the linear combination of the polynomials $F_p^{a,b}[\delta]$,
with coefficients given by the multiplicities with which
the $Sp[\delta](4,{\bf Z})$ orbit $Q_p^{a,b}[\delta]$ appears. 
Thus $P_{01}$ and $P_{02}$ stand for the 
two polynomials corresponding to the two genus 3 orbits of pairs
with 0 common spin structures; $P_{21}, P_{22}, P_{23}$ stand for the 3 orbits 
of pairs with 2 common spin structures; and $P_6$ stands for the single orbit of pairs with 6 common spin structures. The overall sign of
each polynomial is arbitrary. The relative signs are of course fixed
by the stabilizer group of the genus 3 spin structure $\Delta$.
We have (we omit reference to $\Delta$ in $F$),
\bea
P_{01} & = & - 12 F^{0,0} _3 + 4 F^{0,1}_3 + 2 F^{1,1} _6 
\hskip .7in =
-12 \Xi_6 ^2 + 6 F - 2 F^c 
\no \\
P_{02} & = & - 4 F^{0,1} _3 + 4 F^{1,1} _3 + 6 F^{0,0} _6
\hskip .78in =
6 F + 6 F^c
\no \\
P_{21} & = & + 8 F^{0,1} _4 - 2 F^{1,1} _4 + 6 F^{0,0} _6 + 2 F^{1,1} _6
\hskip .2in =
6 F + 12 F^c
\no \\
P_{22} & = & - 4 F ^{0,0} _4 - 2 F^{1,1} _4 + 3 F^{0,0} _6
\hskip .77in 
=4 \Xi _6 ^2 + 3 F + 2 F^c
\no \\
P_{23} & = & - 8 F^{0,0} _4 - 2 F^{1,1} _4 + 2 F^{1,1} _6
\hskip .76in 
= 8 \Xi _6 ^2 + 4 F^c
\no \\
P_6 & = & + 6 F^{0,0} _6 + 4 F^{1,1}_6
\hskip 1.34in 
= 6 F + 4 F^c
\eea
where we have used (\ref{Fexp}) to express all of these in terms of
the quantities $\Xi_6 ^2$, $F$ and $F^c$.
The previous discussion results in the following lemma:

\bigskip
\noindent
{\bf Lemma 4.} {\it Let the sign assignments $\epsilon(\Delta;\{\Delta_i\},\{\Delta_i'\})$ satisfy the transformation (\ref{phase3pairs}) for each orbit ${\cal Q}_{pq}$. Then we have}
\be
\lim_{t\to 0}
\sum_{(\{\Delta_i\},\{\Delta_i'\})\in {\cal Q}_{pq}}
\epsilon(\Delta;\{\Delta_i\},\{\Delta_i'\})
\prod_{i=1}^6
\tet[\Delta_i](0,\Omega^{(3)})^2
\tet[\Delta_i'](0,\Omega^{(3)})^2
=
2^8\eta(\Omega^{(1)})^{24}
P_{pq}(\Omega^{(2)})
\ee

\subsection{Candidates for $\Xi_6[\Delta](\Omega^{(3)})^2$}

Each orbit ${\cal Q}_{pq}$ contributes a consistent term to the candidate for the genus $3$ superstring measure, transforming covariantly under $Sp(6,{\bf Z})$ transformations.
Thus we can take an arbitrary linear combination of these
orbits and obtain a modular covariant expression
\be
\label{combination}
\sum_{p,q}
N_{pq}
\sum_{(\{\Delta_i\},\{\Delta_i'\})\in {\cal Q}_{pq}}
\epsilon(\Delta;\{\Delta_i\},\{\Delta_i'\})
\prod_{i=1}^6
\tet[\Delta_i](0,\Omega^{(3)})^2
\tet[\Delta_i'](0,\Omega^{(3)})^2
\ee
Candidates for $\Xi_6[\Delta](\Omega^{(3)})^2$ must tend to
$2^8\eta(\Omega^{(1)})^2
\,
\Xi_6[\delta](\Omega^{(2)})^2$.
In view of Lemma 4, the limit at $t\to 0$ of the linear combination (\ref{combination}) will be a multiple of $\eta(\Omega^{(1)})^2\,\Xi_6[\delta](\Omega^{(2)})^2$ if the multiplicities $N_{pq}$
satisfy
\bea
\label{N1}
2N_{01}+2N_{02}+2N_{21}+N_{22}+2N_6
&=& 0\nonumber\\
-N_{01}+3N_{02}+6N_{21}+N_{22}+2N_{23}
+2N_6&=& 0
\eea
in which case the limit is given by
\be
\label{N2}
2^8
\
\eta(\Omega^{(1)})^2
\
(-12\,N_{01}+4\,N_{22}+8\,N_{23})
\
\Xi_6[\delta](\Omega^{(2)})^2
\ee
It is convenient to summarize our findings in the following theorem:

\bigskip
\noindent
{\bf Theorem 2.} {\it Let the genus $3$ expression 
$\Xi_6[\Delta](\Omega^{(3)})^2$ be defined by (\ref{Xisquare}),
where $Q_{pq}$ are the orbits of pairs of $\Delta$-admissible sextets from Table.
Assume that the multiplicities $N_{pq}$ satisfy the condition
(\ref{N1}), and set $N=-12\,N_{01}+4\,N_{22}+8\,N_{23}$.
Let $\Delta$ factorize into an even spin structure $\delta$ at genus 2.
Then the expression $\Xi_6[\Delta](\Omega^{(3)})^2$ satisfies
the three conditions}

({\it i'}) $\Xi_6[\Delta](\Omega^{(3)})^2$ {\it is holomorphic on the Siegel upper half space};

\smallskip

({\it ii'}) $\Xi_6[\tilde\Delta](\tilde\Omega^{(3)})^2
=
\det\,(C\Omega^{(3)}+D)^{12}\,\Xi_6[\Delta](\Omega^{(3)})^2$; 

\smallskip

({\it iii'}) 
$\lim _{t\to 0}
\Xi_6[\Delta](\Omega^{(3)})^2
=\eta(\Omega^{(1)})^{24}
\Xi_6[\delta](\Omega^{(2)})^2$.

\bigskip

For example, an integer combination leading to a multiple of $2^8
\
\eta(\Omega^{(1)})^2
\Xi_6[\delta](\Omega^{(2)})^2$
by the square of an integer is $N_{01}=-2$, $N_{02}=4$, $N_{21}=-2$, $N_{23}=-1$, in which case we get
\bea
\label{ans1}
&&
\lim _{t\to 0}
\sum_{(\{\Delta_i\},\{\Delta_i'\})\in {\cal Q}_{pq}}
\epsilon(\Delta;\{\Delta_i\},\{\Delta_i'\})
\prod_{i=1}^6
\tet[\Delta_i](0,\Omega^{(3)})^2
\tet[\Delta_i'](0,\Omega^{(3)})^2
\nonumber\\
&&
\qquad\qquad\qquad
=
16\cdot 2^8\eta(\Omega^{(1)})^{24}
\
\Xi_6[\delta](\Omega^{(2)})^2.
\eea 

\subsection{Vanishing of the genus 3 cosmological constant}

We address a final issue of physical and mathematical significance,
namely the behavior of the genus 3 {\sl cosmological constant}, defined by
\bea
\Upsilon_8  \equiv 
\sum _\Delta \Xi _6 [\Delta] (\Omega ^{(3)}) \tet [\Delta] (0, \Omega ^{(3)})^4
\eea
By the its very construction, $\Xi _6 [\Delta]$ transforms under the 
modular group $Sp (6, {\bf Z})$ as $\tet [\Delta](0,\Omega ^{(3)})^{12}$,
and therefore the quantity $\Upsilon_8$ is a genus 3 modular form of weight 8.
An infinite family of modular forms of weight $4k$ may be generated as follows,
\bea 
\Psi _{4k} (\Omega ^{(3)}) \equiv \sum _{\Delta} 
\tet [\Delta ](0, \Omega ^{(3)})^{8k}
\eea
for $k$ any positive integer. 
In \cite{dp04}, it was argued that $\Psi _8 = \Psi _4 ^2 /8$, based on 
asymptotic identifications and numerical calculations.
We shall assume that this is the only independent holomorphic modular form 
of weight 8, as we are not aware of any proof that this statement is true.
Given this assumption, as well as the asymptotic behavior established 
in this paper for $\Xi _6 [\Delta] (\Omega ^{(3)}) $, as the surface undergoes 
a separating degeneration, it is clear that the modular form  $\Upsilon_8$
must vanish in this limit. But $\Psi _8$ is non-zero in the same limit.
As a result, $\Upsilon _8 =0$ throughout moduli space, and the 
cosmological constant vanishes to three loop order.

\bigskip

\noindent
{\large \bf Acknowledgments}

\medskip

We are happy to thank Edward Witten for stimulating discussions. 
Part of this work was carried out while one of us (E.D.) was at
the Aspen Center for Physics. All calculations were carried out using Maple 9.

\end{document}